\documentclass[twocolumn]{aastex62}
\usepackage{lineno}
\usepackage{tabularx}
\usepackage{amsmath,amssymb}
\usepackage{color}
\usepackage{xcolor}
\usepackage{multirow}
\usepackage{CJK}
\usepackage{threeparttablex}
\usepackage{extarrows}
\definecolor{iceberg}{rgb}{0.44, 0.65, 0.82}
\definecolor{lavenderblue}{rgb}{0.8, 0.8, 1.0}
\definecolor{lavenderpink}{rgb}{0.98, 0.68, 0.82}
\usepackage{subfigure}

\submitjournal{ApJ}
\graphicspath{{./}{figures/}}
\usepackage{color, colortbl}
\usepackage{romannum}
\begin{document}
\begin{CJK*}{UTF8}{gbsn}

\title{Temporal and Spectral Analysis of the Unique and Second Brightest Gamma-Ray Burst GRB 230307A: Insights from GECAM and \textit{Fermi}/GBM Observations}

\author[0000-0002-2516-5894]{R.~Moradi}
\affiliation{Key Laboratory of Particle Astrophysics, Institute of High Energy Physics, Chinese Academy of Sciences, Beijing 100049, People’s Republic of China}

\author{C.~W.~Wang}
\affiliation{Key Laboratory of Particle Astrophysics, Institute of High Energy Physics, Chinese Academy of Sciences, Beijing 100049, People’s Republic of China}
\affiliation{University of Chinese Academy of Sciences, Chinese Academy of Sciences, Beijing 100049, China}

\author[0000-0002-9725-2524]{B.~Zhang (张冰)}
\affiliation{Nevada Center for Astrophysics, University of Nevada Las Vegas, NV 89154, USA}
\affiliation{Department of Physics and Astronomy, University of Nevada Las Vegas, NV 89154, USA}

\author[0000-0001-7959-3387]{Y.~Wang (王瑜)}
\affiliation{ICRANet, Piazza della Repubblica 10, I-65122 Pescara, Italy}
\affiliation{ICRA, Dipartamento di Fisica, Sapienza Universit\`a  di Roma, Piazzale Aldo Moro 5, I-00185 Rome, Italy}
\affiliation{INAF, Osservatorio Astronomico d'Abruzzo, Via M. Maggini snc, I-64100, Teramo, Italy} 

\author{S.-L. Xiong (熊少林)}
\affiliation{Key Laboratory of Particle Astrophysics, Institute of High Energy Physics, Chinese Academy of Sciences, Beijing 100049, People’s Republic of China}

\author{S.-X. Yi}
\affiliation{Key Laboratory of Particle Astrophysics, Institute of High Energy Physics, Chinese Academy of Sciences, Beijing 100049, People’s Republic of China}

\author{W.-J. Tan}
\affiliation{Key Laboratory of Particle Astrophysics, Institute of High Energy Physics, Chinese Academy of Sciences, Beijing 100049, People’s Republic of China}
\affiliation{University of Chinese Academy of Sciences, Chinese Academy of Sciences, Beijing 100049, China}

\author{M. Karlica} 
\affiliation{Astronomical observatory Belgrade, Volgina 7, 11060 Belgrade, Serbia}

\author{S.-N. Zhang (张双南)}  
\affiliation{Key Laboratory of Particle Astrophysics, Institute of High Energy Physics, Chinese Academy of Sciences, Beijing 100049, People’s Republic of China}
\affiliation{University of Chinese Academy of Sciences, Chinese Academy of Sciences, Beijing 100049, China}

\email{rmoradi@ihep.ac.cn;cwwang@ihep.ac.cn;bing.zhang@unlv.edu;\\yu.wang@inaf.it;xiongsl@ihep.ac.cn}

\date{Received date /Accepted date }

\begin{abstract}
In this study, we present the pulse profile of the unique and the second brightest gamma-ray burst GRB 230307A, and analyze its temporal behavior using a joint GECAM--\textit{Fermi}/GBM time-resolved spectral analysis. The utilization of GECAM data is advantageous as it successfully captured significant data during the pile-up period of the \textit{Fermi}/GBM. We investigate the evolution of its flux, photon fluence, photon flux, peak energy, and the corresponding hardness-intensity and hardness-flux correlations. The findings within the first 27 seconds exhibit consistent patterns reported previously, providing valuable insights for comparing observations with predictions from the synchrotron radiation model invoking an expanding shell. Beyond the initial 27 seconds, we observe a notable transition in the emitted radiation, attributed to high latitude emission (HLE), influenced by the geometric properties of the shells and the relativistic Doppler effects. By modeling the data within the framework of the large-radius internal shock model, we discuss the required parameters as well as the limitations of the model. We conclude that a more complicated synchrotron emission model is needed to fully describe the observational data of GRB 230307A.

\end{abstract}

%%%%%%%%%%%%%%%%%%%%%%%%%%%%%%%%%%%%%%%%%%%%%%%%%%

%%%%%%%%%%%%%%%%% BODY OF PAPER %%%%%%%%%%%%%%%%%%

%%%%%%%%%%%%%%%%%%%%%%%%%%%%%%%%%%%%%%%%%%%%%%%%%%%%
%%%%%%%%%%%%%%%%%%%%%%%%%%%%%%%%%%%%%%%%%%%%%%%%%%%%
\section{Introduction}\label{sec:1}
%%%%%%%%%%%%%%%%%%%%%%%%%%%%%%%%%%%%%%%%%%%%%%%%%%%%
%%%%%%%%%%%%%%%%%%%%%%%%%%%%%%%%%%%%%%%%%%%%%%%%%%%%

A comprehensive understanding of the physical origins of gamma-ray burst (GRB) pulses is crucial to unravel the complexities of the overall GRB phenomenon and to shed light on the nature of the inner engine and associated radiation processes \citep[see e.g.][]{1996ApJ...459..393N}.

The individual GRB pulses typically exhibit a characteristic temporal evolution, characterised by a rapid rise followed by an exponential decay, often referred to as the ``FRED'' shape \citep{1994ApJS...92..229F,zhang2018physics}.

However, statistical investigations of the decay phase of smooth FRED pulses within the Burst and Transient Source Experiment (BATSE) have shown that many of these pulses do not conform to exponential behaviour \citep{1996AIPC..384...96S,2000ApJ...529L..13R,2002ApJ...566..210R}. Therefore, the exploration of alternative functions, such as power-law functions, has been proposed to explain this decay trend \citep{2000ApJ...529L..13R,2002ApJ...566..210R}.

During the decay phase, the spectral hardness of GRB pulses demonstrates a noticeable decrease. A significant fraction of pulses in this phase appear to satisfy specific relations between their temporal and spectral properties, such as the hardness-intensity correlation (HIC) \citep{1983Natur.306..451G} and the hardness-fluence correlation (HFC) \citep{1996AIPC..384..202L}.

Based on these empirical findings, and with the goal of comprehending the radiative mechanisms responsible for the cooling process, it is essential to concentrate on the decay phase of GRBs' prompt emission. Furthermore, it is vital for theoretical models to replicate these observational results, as they can help distinguish among various possibilities; \citep[see e.g.,][]{2010MNRAS.405..695G, 2014IJMPD..2330002Z, 2021A&A...656A.134G, 2022Univ....8..310P}. 

In this regard, in this paper we conduct a comprehensive investigation into the spectral and temporal characteristics of GRB 230307A, the second brightest gamma-ray burst (GRB) detected in over 50 years, characterized by its long-duration prompt emission \citep{2023arXiv231201074D, 2023ApJ...954L..29D}. The unique features of GRB 230307A, along with the similar properties of GRB 211211A, which exhibited a long gamma-ray followed by softer temporally extended emission (EE) and kilonova \textcolor{black}{\citep{2023arXiv230800633G, 2023arXiv230702098L,308e4fe1b362453380598f84c04a1f83}}, suggest a new GRB class associated with the compact stellar merger origin of GRB 230307A \citep{2023arXiv231201074D, 2023ApJ...958L..33G,2024arXiv240702376W}. The emergence of an X-ray component after the prompt $\gamma$-ray emission phase suggests a magnetar engine \citep{2023arXiv230705689S} (see also \cite{2023ApJ...954L..29D}). Detailed analyses of the temporal spectral behavior suggest that the emission is likely synchrotron radiation from a Poynting-flux-dominated jet, with evidence of mini-jets suggested in the internal-collision-induced magnetic reconnection and turbulence (ICMART) model \citep{2023arXiv231007205Y}, even though other possibilities have also been discussed \citep[e.g.][]{2023ApJ...953L...8W}. 

The high brightness of GRB 230307A caused the Fermi Gamma-ray Burst Monitor \citep[\textit{Fermi}/GBM;][]{2023GCN.33411....1D} to encounter pulse pile-up and data loss during the prompt emission from $T_0$+1 s to $T_0$+20 s \citep{2023GCN.33551....1D}. This issue was also observed in the case of another bright GRBs, e.g. GRB 221009A \citep{2023ApJ...952L..42L}. However, the Gravitational wave high-energy Electromagnetic Counterpart All-sky Monitor \citep[GECAM;][]{2023GCN.33406....1X}, successfully obtained significant data during the pile-up period of the \textit{Fermi}/GBM, showcasing its unique capability to capture data from highly luminous GRBs. This ability of GECAM provides valuable insights into the nature of exceptionally bright GRBs.

Utilizing data obtained from both GECAM and \textit{Fermi}/GBM, we perform a time-resolved spectral analysis during the first 100 seconds of GRB 230307A. Our analysis, as shown in section~\ref{sec:2} reveals consistent behavior with the Hardness-Intensity Correlation (HIC) and Hardness-Fluence Correlation (HFC) relations as reported by \cite{2000ApJ...529L..13R,2002ApJ...566..210R} derived from a sample of 25 bursts.

In view that this spectral evolution behavior is consistent with a general synchrotron radiation model involving a large emission radius, we investigate how the model may be consistent with such models, especially the widely discussed internal shock model \citep{1994ApJ...430L..93R,1996AIPC..384..782S,1997MNRAS.287..110S,1997ApJ...490...92K, 2001grba.conf...87S}. 

We consider relativistic ejection as a continuous process, with internal shocks taking the form of propagating shock waves within the outflow. We utilize a version of the internal shock model developed by \cite{1998MNRAS.296..275D, 2003MNRAS.342..587D}, which effectively discusses one pair of internal shocks at a large emission radius. This approach results in pulse shapes during the decay phase, determined by the hydrodynamical timescale associated with the propagation of these shock waves. 

We point out the limitations of this model and discuss possible solutions to improve the model. In particular, we echo the suggestion of \cite{2023arXiv231007205Y} that the full dataset of GRB 230307A demands a more complicated large-radius synchrotron radiation model invoking mini-jets, as envisaged within the framework of ICMART model \citep{2011ApJ...726...90Z}. 

The paper is organized as follows:
Section~\ref{sec:2} presents the detailed methodology of the time-resolved spectral analysis applied to GRB 230307A using data from GECAM and Fermi-GBM.
In section~\ref{sec:3}, we provide a brief overview of the internal shocks (IS) model employed in this study.
Section~\ref{sec:4} focuses on investigating the astrophysical model parameters required in the IS model to explain the observed temporal and spectral evolution of GRB 230307A and the limitations of the model. Finally, in section~\ref{sec:5}, we present the concluding remarks of our study.

%%%%%%%%%%%%%%%%%%%%%%%%%%%%%%%%%%%%%%%%%%%%%%%%%%%%
%%%%%%%%%%%%%%%%%%%%%%%%%%%%%%%%%%%%%%%%%%%%%%%%%%%%
\section{Observation of GRB 230307A}\label{sec:2}
%%%%%%%%%%%%%%%%%%%%%%%%%%%%%%%%%%%%%%%%%%%%%%%%%%%%
%%%%%%%%%%%%%%%%%%%%%%%%%%%%%%%%%%%%%%%%%%%%%%%%%%%%  
On March 7, 2023, at 15:44:06 UT ($T_0$) the Gravitational wave high-energy Electromagnetic Counterpart All-sky Monitor (GECAM) was triggered in-flight by this exceptionally bright long burst and the low latency alert reporting the GECAM discovery of this extremely bright burst initiated a global observation campaign \citep{2023GCN.33406....1X}. The GECAM light curve shows a roughly FRED shape with a possible precursor and an overall duration of about 100 seconds \citep{2023GCN.33406....1X}. The GECAM duration (T90) of this burst is 41.52 $\pm$ 0.03 s in the 10 -- 1000 keV energy range \citep{2023arXiv230705689S}. At the same time, the Fermi Gamma-ray Burst Monitor (GBM) detected and located GRB 230307A. The GBM light curve of this event exhibited a single burst in 10-1000 keV with a duration (T90) of approximately 35 seconds \citep{2023GCN.33411....1D}. With a redshift estimated to be $z\sim$0.065 \citep{2023GCN.33485....1G}, the isotropic energy measured by GECAM-B is $E_{\rm iso}$ =  4.8$\times$10$^{52}$~erg.

%%%%%%%%%%%%%%%%%%%%%%%%%%%%%%%%%%%%%%%%%%%%%%%%%%%%%%%%
\subsection{Details of GECAM and \textit{Fermi}/GBM  observations}
%%%%%%%%%%%%%%%%%%%%%%%%%%%%%%%%%%%%%%%%%%%%%%%%%%%%%%%%

%%%%%%%%%%%%%%%%%%%%%%%%%%%%%%%%%%%%%%%%%%%%%%%%%%%%%%%%
\subsubsection{GECAM}
%%%%%%%%%%%%%%%%%%%%%%%%%%%%%%%%%%%%%%%%%%%%%%%%%%%%%%%%

As of the detection of GRB 230707A, GECAM consists of three separate satellites: GECAM-A, GECAM-B, and GECAM-C. GECAM-A and GECAM-B were launched in December 2020 \citep{Li2021RDTM}, while GECAM-C,  onboard the SATech-01 satellite, was launched in July 2022 \citep{ZHANG2023168586}.

Each GECAM satellite has two kinds of scientific instruments: gamma-ray detectors (GRDs) and charged particle detectors (CPDs). There are 25 GRDs onboard GECAM-B and 12 GRDs onboard GECAM-C. GRB 230307A was detected by both GECAM-B and GECAM-C, and neither GECAM-B nor GECAM-C suffered saturation during the event. GRD04 of GECAM-B and GRD01 of GECAM-C are selected for spectral analysis due to their smallest zenith angles to the burst. These two detectors operate in two readout channels: high gain (HG) and low gain (LG), which are independent in terms of data processing, transmission, and dead time. 

For GRD04 of GECAM-B, the energy range of the HG channel data is used from about 40 keV to 300 keV, while the energy range of the LG channel data is used from about 700 keV to 8000 keV. For GRD01 of GECAM-C, only the HG channel data is used with an energy range from 6 keV to 100 keV. It should be noted that the response of GRD01 of GECAM-C for the 6-15 keV range is affected by certain instrumental effects, which may result in inaccuracies. This issue is currently being studied. The background of GECAM-B is estimated by fitting the data from $T_0$-50\,s to $T_0$-5\,s and $T_0$+160\,s to $T_0$+200\,s with first order polynomials. 

The background of GECAM-C is estimated by fitting the data from $T_0$-20\,s to $T_0$-1\,s and $T_0$+170\,s to $T_0$+600\,s with a combination of first and second-order exponential polynomials. 

\textcolor{black}{It is noteworthy that GECAM features a unique and optimized design specifically for detecting extremely bright GRBs. In addition to employing transmission channels with large capacity and broad bandwidth, adjacent detectors with similar pointing are read out by independent electronic modules. Thanks to these sophisticated designs, the count rate upper limit of each GRD can reach up to 100 kcps, significantly exceeding the count rate of GRB 230307A detected by GECAM.}

%%%%%%%%%%%%%%%%%%%%%%%%%%%%%%%%%%%%%%%%%%%%%%%%%%%%%%%%
\subsubsection{\textit{Fermi}/GBM }
%%%%%%%%%%%%%%%%%%%%%%%%%%%%%%%%%%%%%%%%%%%%%%%%%%%%%%%%

In this study, we also utilize the Time-Tagged Event (TTE) data from the \textit{Fermi}/GBM instrument, which is one of the two scientific instruments on board the \textit{Fermi} satellite, the other being \textit{Fermi}/LAT. The \textit{Fermi}/GBM instrument is equipped with two types of detectors, namely 12 Sodium Iodide (NaI) detectors (n0-nb) and 2 Bismuth Germanate (BGO) detectors (b0-b1). 

Due to the high brightness of GRB 230307A, GBM experienced pulse pile-up and data loss during the prompt emission, resulting in the creation of Bad Time Intervals (BTI) as defined in \citep{2023GCN.33551....1D}. To supplement the GECAM data when the source is not as bright (prior to  $T_0$+0.65\,s and after $T_0$+17.5\,s), we carry out a joint analysis using GBM data, using the na detector ranging from 8 keV to 900 keV, and the b1 ranging from 0.3 MeV to 8 MeV. 

%%%%%%%%%%%%%%%%%%%%%%%%%%%%%%%%%%%%%%%%%%%%%%%%%%%%%%%%
\subsection{Time--resolved spectral analysis }
%%%%%%%%%%%%%%%%%%%%%%%%%%%%%%%%%%%%%%%%%%%%%%%%%%%%%%%%
The spectral analysis is conducted using the Pyxspec software \citep{2021ascl.soft01014G}. The exceptional brightness of GRB 230307A enables the execution of spectral analysis with a significantly high time resolution. The time intervals used in this study are determined based on the results of the Bayesian block analysis \citep{2013ApJ...764..167S}, with additional adjustments made based on the fitting results. 

Specifically, during the period of high brightness from $T_0$ to $T_0$+15\,s, the time intervals are set to 0.1\,s in order to achieve a more refined time-resolved spectrum. During the period of decreased burst intensity, specifically from $T_0$+ 15\,s to $T_0$+ 32\,s, the time intervals are adjusted to 0.5\,s. This choice allows for a more refined analysis of the dip spectrum, ensuring an adequate number of photons for accurate fitting. Following $T_0$+32 s, the time intervals are primarily determined based on the Bayesian blocks analysis results. However, these intervals are further adjusted multiple times to ensure optimal fitting of each bin, aiming for narrow and well-fitted bins.

Three distinct photon spectral models are chosen for analysis: BAND model \citep{1993ApJ...413..281B}, the power law with a high-energy exponential cutoff (CPL) model, and the simple power law (PL) model. In addition, we incorporate the uncertainties arising from cross-calibration by introducing multiplicative factors between different detectors. Table \ref{tab:fit_result} presents the best-fit parameters obtained for each time interval of GRB 230307A. \textcolor{black}{The columns include the time interval, the best-fit model, the $\alpha$ parameter for the PL, CPL, or BAND function when applicable, the $\beta$ parameter for the BAND function in intervals where it is the best fit, the peak energy, the flux, the photon flux, the CSTAT value, and the degrees of freedom (DOF) to assess the goodness of fit for the best model. Additionally, we thoroughly explored the potential need for an additional spectral component, such as a thermal component, to fit the spectra. However, we did not find any statistically significant evidence supporting the necessity of such an extra component.}

The optimal model is determined using the Bayesian Information Criterion \citep[BIC,][]{1978AnSta...6..461S}. The BIC is defined as ${\rm BIC} = -2{\rm ln} L + k{\rm ln} N$, where $L$ represents the maximum likelihood value, $k$ denotes the number of free parameters in the model, and $N$ signifies the number of data points. A model with a lower BIC value is preferred, particularly when the difference in BIC ($\Delta_{\rm BIC}$) exceeds 10. In the best model, all parameters should be well--constrained, and the multiplicative factors should be approximately equal to 1. 

\begin{figure*}
\centering

\includegraphics[width=0.92\hsize,clip]{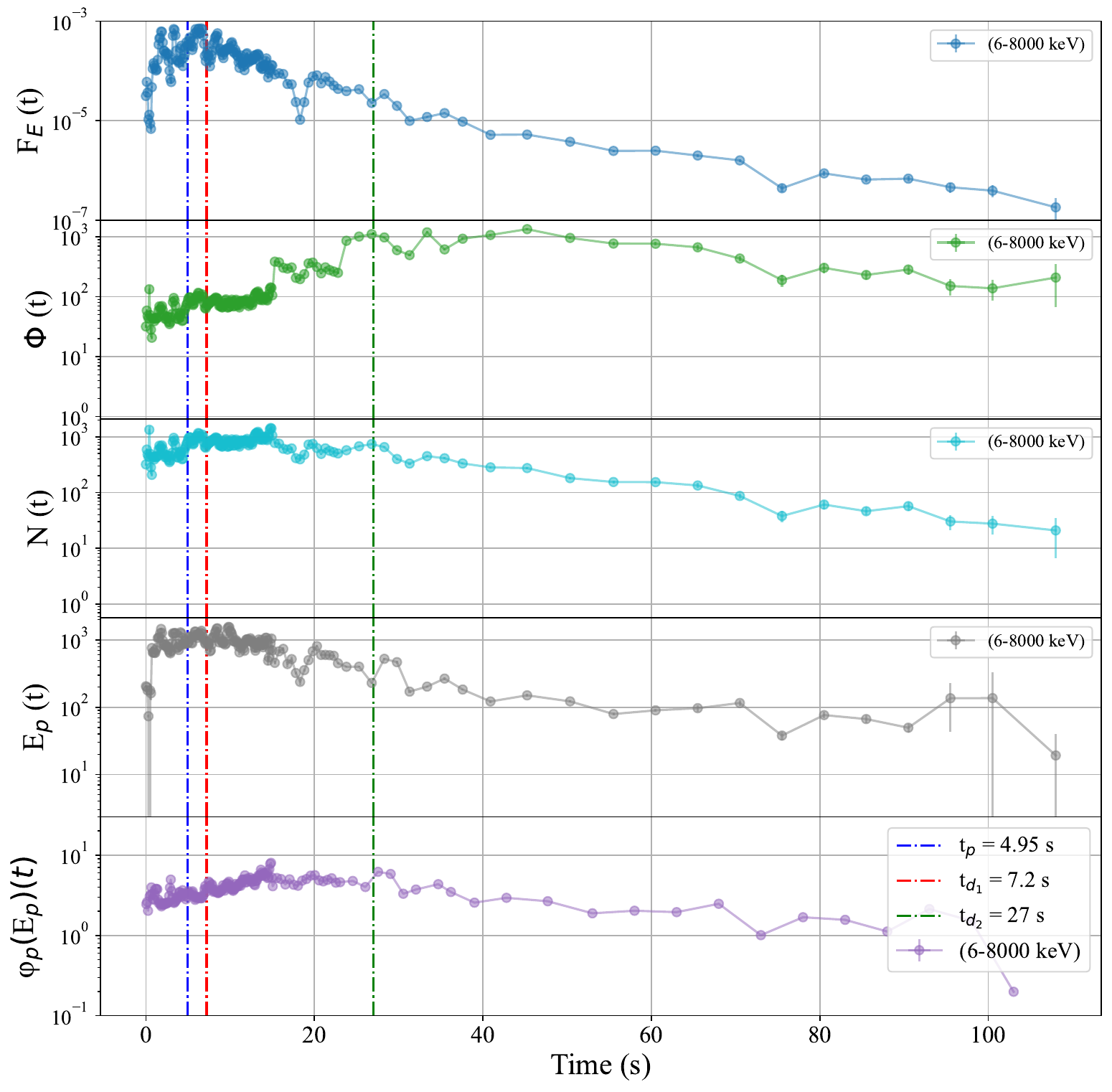} 

\caption{Joint GECAM+Fermi/GBM temporal and spectral Evolution of GRB 230307A. The first panel displays the flux evolution, $F_{ E} (t)$, in erg.cm$ ^{-2}$.s $^{-1}$. The second panel shows the photon fluence evolution, $\mathrm{\Phi} (t)$, in photons.cm$ ^{-2}$. The third panel shows the photon flux evolution, $N(t)$, in photons.cm$ ^{-2}$.s $^{-1}$. The forth panel shows the spectral peak energy evolution, $E_{\rm p} (t)$, in keV. The fifth panel shows $ \varphi_{\rm p} (E_{\rm p})$ defined in Eq.~\ref{eq:FE}. The peak of the pulse is $t_{\rm p}$ = 4.95 s. The interval from $t_{d_1}$ = 7.2 s to $t_{d_2}$ = 27 s  indicates the time interval of the decay phase studied.}, \label{fig:Model:M} 
\end{figure*}

\begin{figure*}
\centering

\includegraphics[width=0.92\hsize,clip]{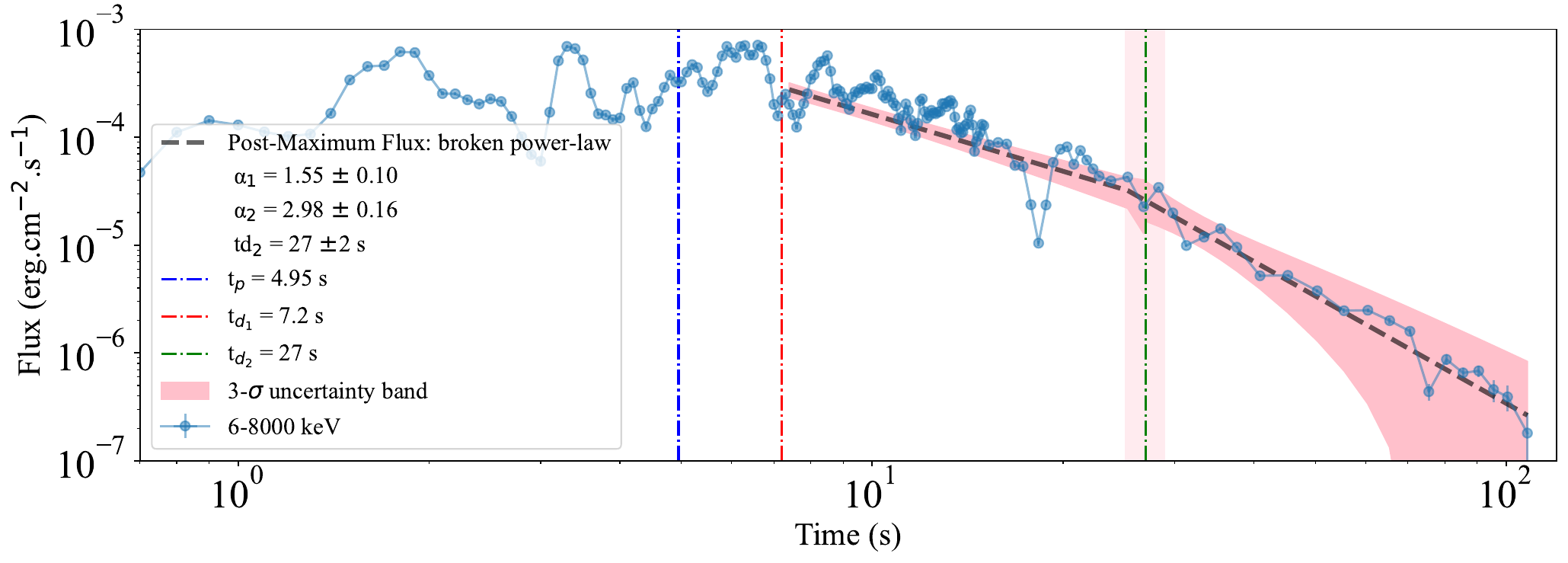} 
\caption{The post maximum flux $F(t)$ best--fitted with a broken power-law model. The best fit parameters indicate a break time at 27$\pm$2 s, with the first index $\alpha_1 = 1.55 \pm 0.10 $ and the second index $\alpha_2 = 2.98 \pm 0.16$. 
}, \label{fig:BPL} 
\end{figure*}

\begin{figure*}
\centering

\textbf{A}\includegraphics[width=0.35\hsize,clip]{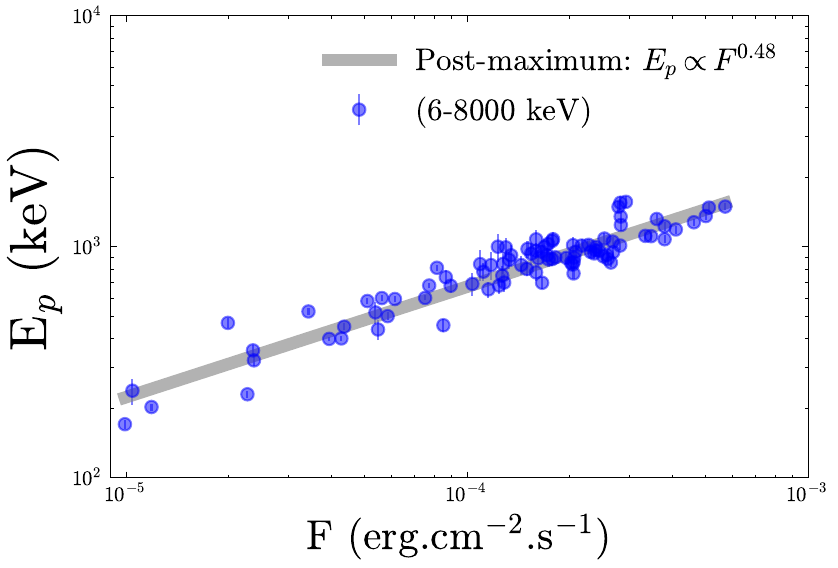} 
\includegraphics[width=0.35\hsize,clip]{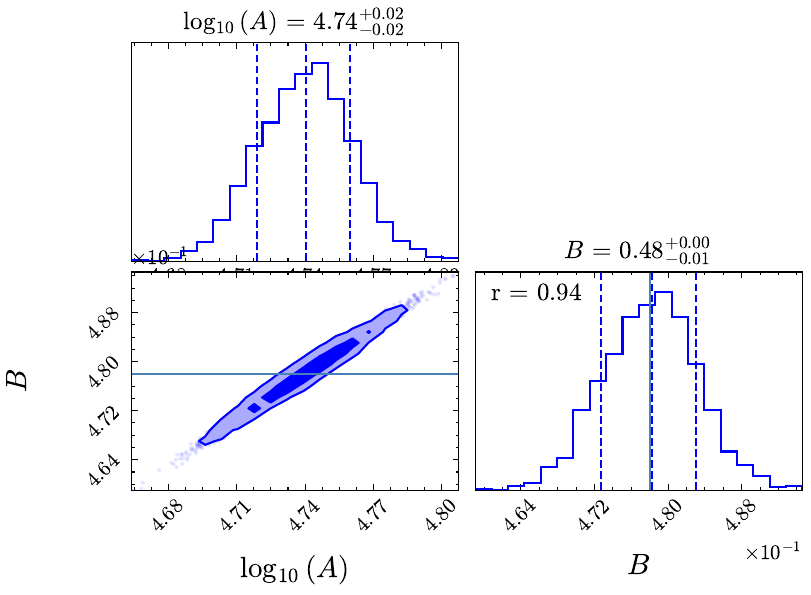}
\textbf{B}\includegraphics[width=0.35\hsize,clip]{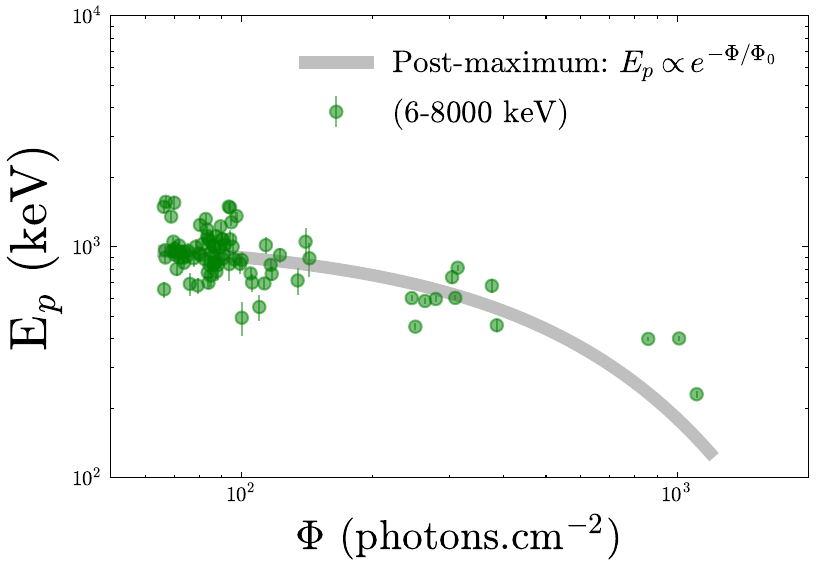} 
\includegraphics[width=0.35\hsize,clip]{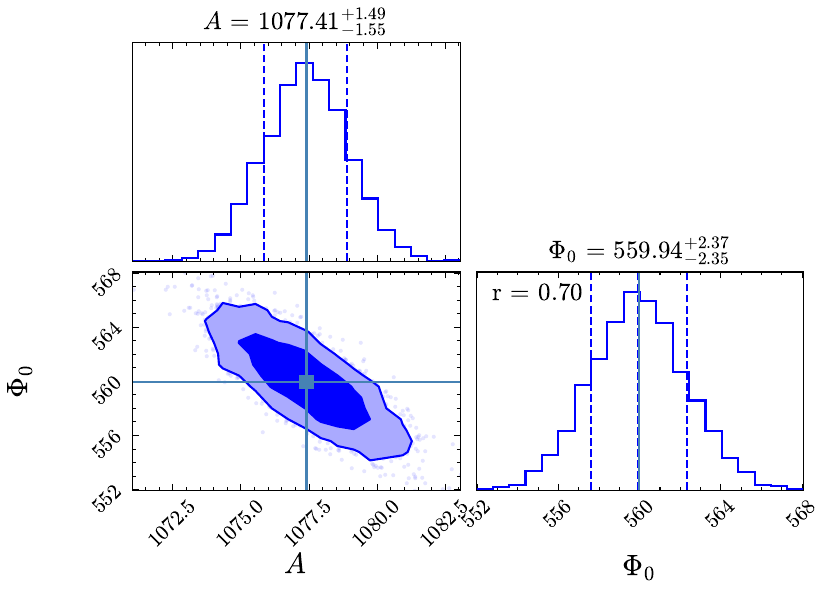}
\textbf{C}\includegraphics[width=0.35\hsize,clip]{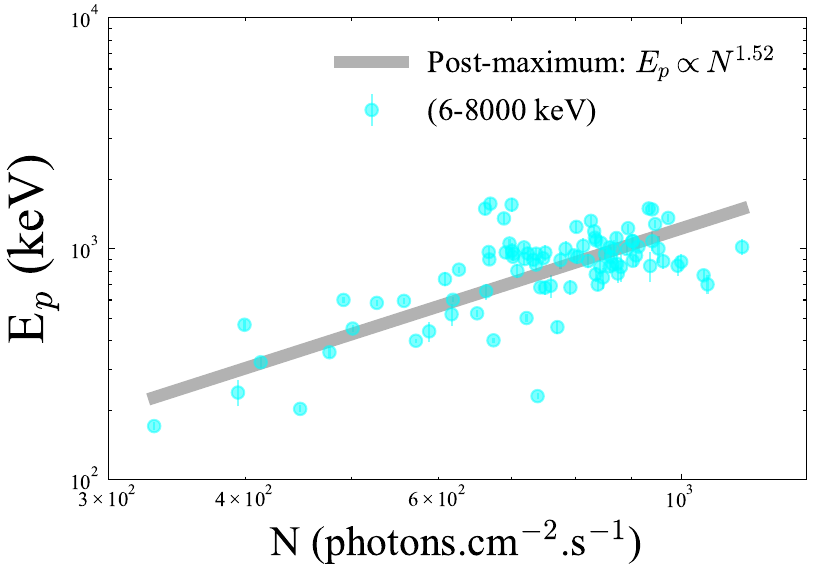} 
\includegraphics[width=0.35\hsize,clip]{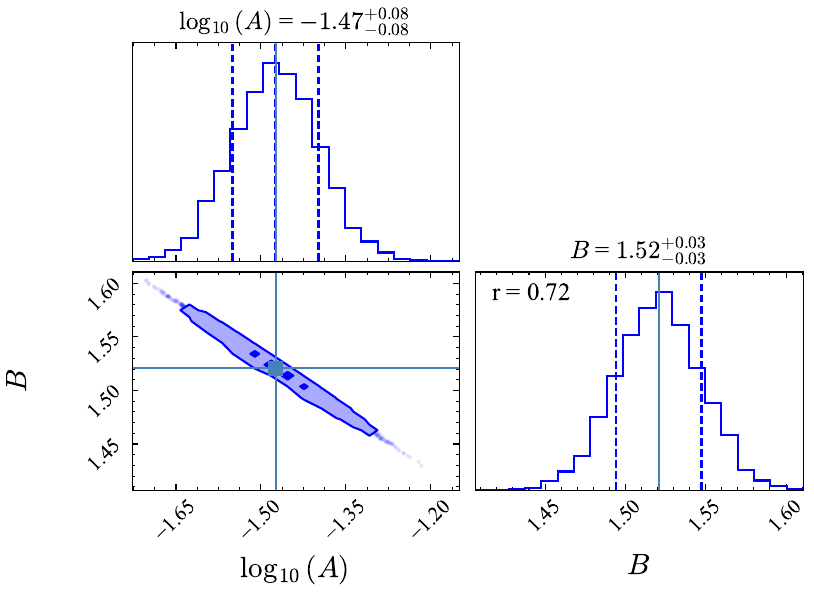}
\textbf{D}\includegraphics[width=0.35\hsize,clip]{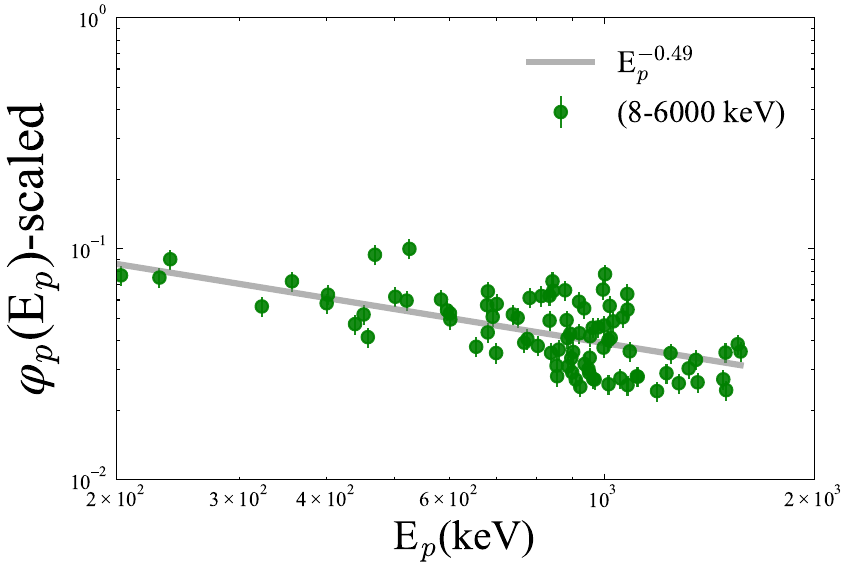} 
\includegraphics[width=0.35\hsize,clip]{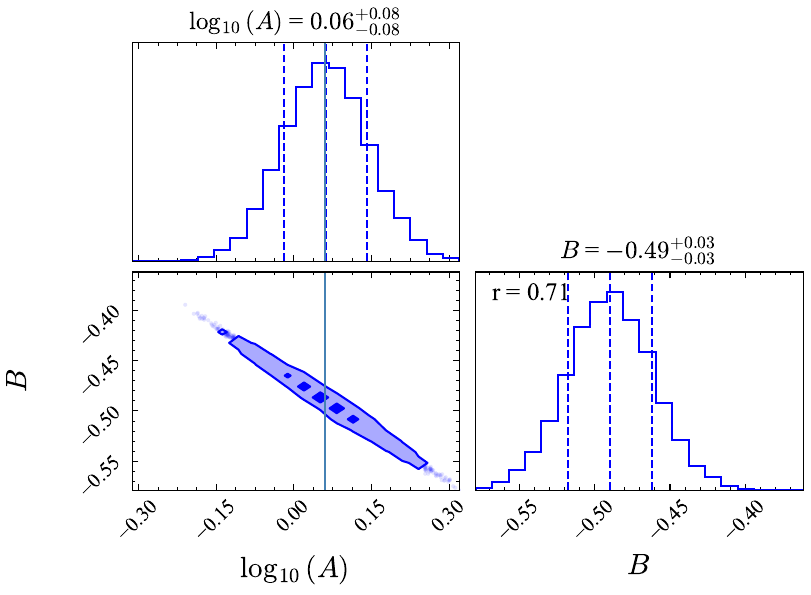}
\caption{\textcolor{black}{The correlations between the peak energy, $E_\mathrm{p}(t)$, and various parameters such as \textbf{A:} the flux, $F_{E}(t)$, $N(t)$, \textbf{B:} photon fluence,  \textbf{C:} photon number, $\mathrm{\Phi}(t)$, and \textbf{D:} $ \varphi_{\rm p} (E_{\rm p})$, are explored in our analysis. The relationship between $E_\mathrm{p}(t)$ and $F_{E}(t)$ aligns with the findings of previous studies by \cite{2004ApJ...606L..29L}, \cite{2009A&A...496..585G}, and \cite{2012ApJ...756..112L}. The Hardness-Fluence Correlation (HFC) is illustrated panel \textbf{B}, while the Hardness-Intensity Correlation (HIC) is shown in the panel \textbf{C}. The optimal fitting time interval ranges from $t_{d_1} = 7.2$ to $t_{d_2} = 27$ seconds. The correlation coefficients of 0.94, 0.70, 0.72, and 0.71 demonstrate strong relationships between $E_\mathrm{p}(t)$ and $F_{E}(t)$, $\mathrm{\Phi}(t)$, $N(t)$, and $\varphi_{\rm p}(E_{\rm p})$, respectively. The fits for $E_\mathrm{p}(t)$ with $F_{E}(t)$, $N(t)$, and $\varphi_{\rm p}(E_{\rm p})$ are performed using a logarithmic model: $\log(E_{\rm p}) = A + B \cdot \log(x)$, where $x$ represents $F_{E}(t)$, $N(t)$, and $\varphi_{\rm p}(E_{\rm p})$.}}\label{fig:Model:N} 
\end{figure*}

%%%%%%%%%%%%%%%%%%%%%%%%%%%%%%%%%%%%%%%%%%%%%%%%%%%%
\subsection{GECAM+Fermi/GBM Temporal and Spectral Evolution}
%%%%%%%%%%%%%%%%%%%%%%%%%%%%%%%%%%%%%%%%%%%%%%%%%%%%

The characteristics of the gamma-ray of the burst can be inferred from the temporal changes in the light curve and the spectral properties; \citep[see e.g.,][]{2007ApJ...671.1903C, 2021PhRvD.104f3043M, 2021MNRAS.504.5301R, 2021A&A...656A.134G}. These observations have uncovered various statistical relationships between different measurable quantities, providing valuable insights and limitations on the physical models that govern the production of gamma-rays \citep[see e.g.,][and references therein]{2023ApJ...949..110T}. The advent of new satellites like GECAM \citep{2021arXiv211204775Z} has further enhanced our ability to study and understand these phenomena.

As explained in the previous subsections, we examined the potential evolution of the spectral shape during the decay phase of GRB 230307A, but no significant changes were observed; see also Table.~\ref{tab:fit_result}. We now study the temporal behaviour of the energy flux $F_\mathrm{E}(t)$, the photon flux $N(t)$ and spectral peak energy $E_\mathrm{p}(t)$. $F_\mathrm{E}(t)$ is related to $N(t)$ and $E_\mathrm{p}(t)$ by \citep{2003MNRAS.342..587D}

\begin{equation}
F_\mathrm{E}(t)=N(t)E_{\rm p}(t)\;\frac{\varphi_0}{\varphi(E_\mathrm{p})}
\label{eq:FE}
\end{equation}
with
\begin{equation}
\varphi(E_\mathrm{p})=\int_{E_1/E_\mathrm{p}}^{E_2/E_\mathrm{p}}
\mathcal{N}(x)\mathrm{d}x, ~~~~ \varphi_0 = \int_0^{\infty} x \mathcal{N}(x)\mathrm{d}x, \ \nonumber 
\end{equation}
with $\mathcal{N}(x)$ representing the spectral shape. Figure.~\ref{fig:Model:M} shows the joint GECAM+Fermi/GBM evolution of the above functions, namely $F_\mathrm{E}(t)$, $N(t)$, $E_\mathrm{p}(t)$, $\varphi(E_\mathrm{p}) (t)$, as well as the photon fluence $\mathrm{\Phi} (t)$. These functions are obtained from the best fit parameters determined in Section \ref{sec:2}, and reported in Table.~\ref{tab:fit_result}.

\citet{2002ApJ...566..210R} have shown that the  decay behavior of the photon flux, $N(t)$, is a consequence of the validity of both the Hardness-Intensity Correlation (HIC) 
\begin{equation}
E_\mathrm{p}(t)\propto N(t)^{\delta}
\label{eq:HIC}
\end{equation}
and the Hardness-Fluence Correlation (HFC), 
\begin{equation}
E_\mathrm{p}(t)\propto e^{-\Phi_{\rm N}(t)/\Phi_0}\ ,
\label{eq:HFC}
\end{equation}
where $\Phi_0$ is an exponential decay constant \citep{2000ApJ...529L..13R}, and $N(t)$ follows a generalized power law during the decay phase:
\begin{equation}
N(t) = \frac{N_{0}}{(1+t/\tau)^{n}}\ ,
\label{eq:Nbis}
\end{equation}
where  $t=0$ represents the starting time of the decay phase, and it serves as a reference point from which the decay behavior of the photon flux $N(t)=N_0$ is measured, and $\tau$ is a constant. Subsequently, \citet{2002ApJ...566..210R} found that the distribution of $n$ in their sample was peaked around $n=1$.

Based on our comprehensive analysis of data obtained from the joint analysis of GECAM and Fermi/GBM, we find that none of $F_\mathrm{E}(t)$, $E_\mathrm{p}(t)$, and $N(t)$ can be accurately described by a single decaying power law over the entire post maximum emission period, extending from the initial decay time $t_{\rm d_1}$ = 7.2 s to $t\sim$ 100 s. 

We then investigate the behavior of the flux and fit it using a broken power-law model. The best fit parameters, as shown in Fig.~\ref{fig:BPL} reveals a break time $t = 27\pm$2 s, with the first index $\alpha_1 = 1.55 \pm 0.10 $ and the second index $\alpha_2 = 2.98 \pm 0.16$. Therefore, we denote the break time of the flux as $t_{\rm d_2}$ = 27 s, representing the initiation of the second decay phase of the flux. 

\begin{figure*}
\centering

\includegraphics[width=0.99\hsize,clip]{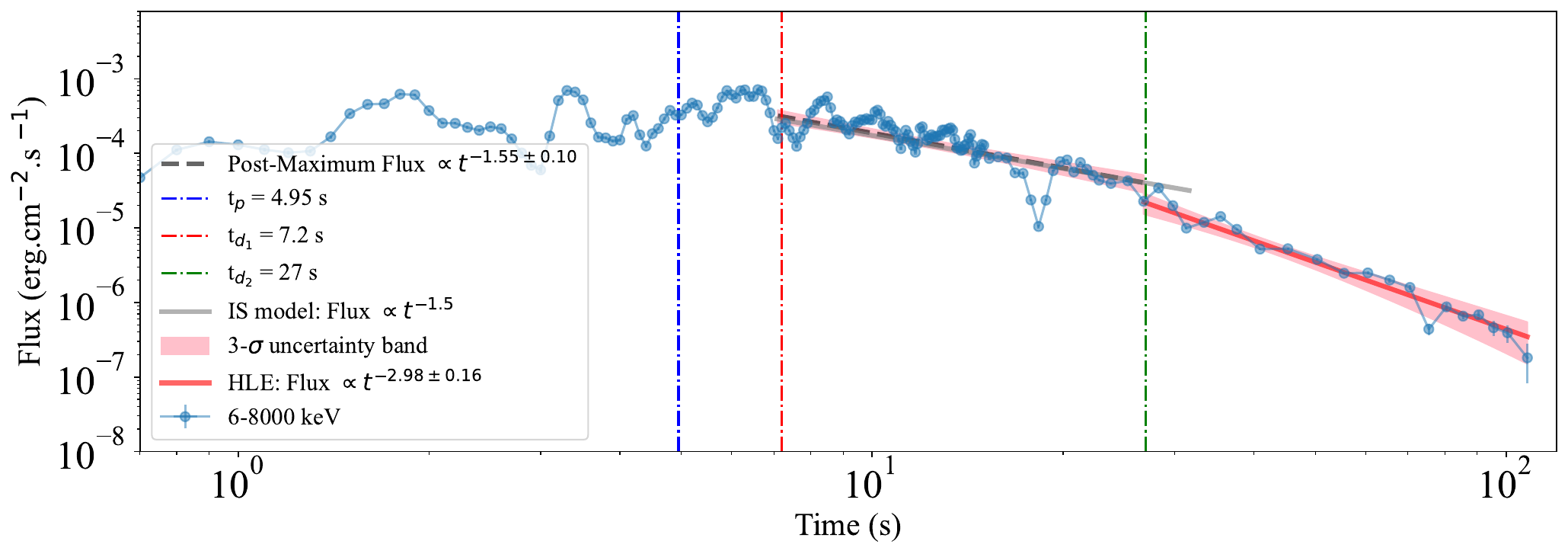} 
\includegraphics[width=0.99\hsize,clip]{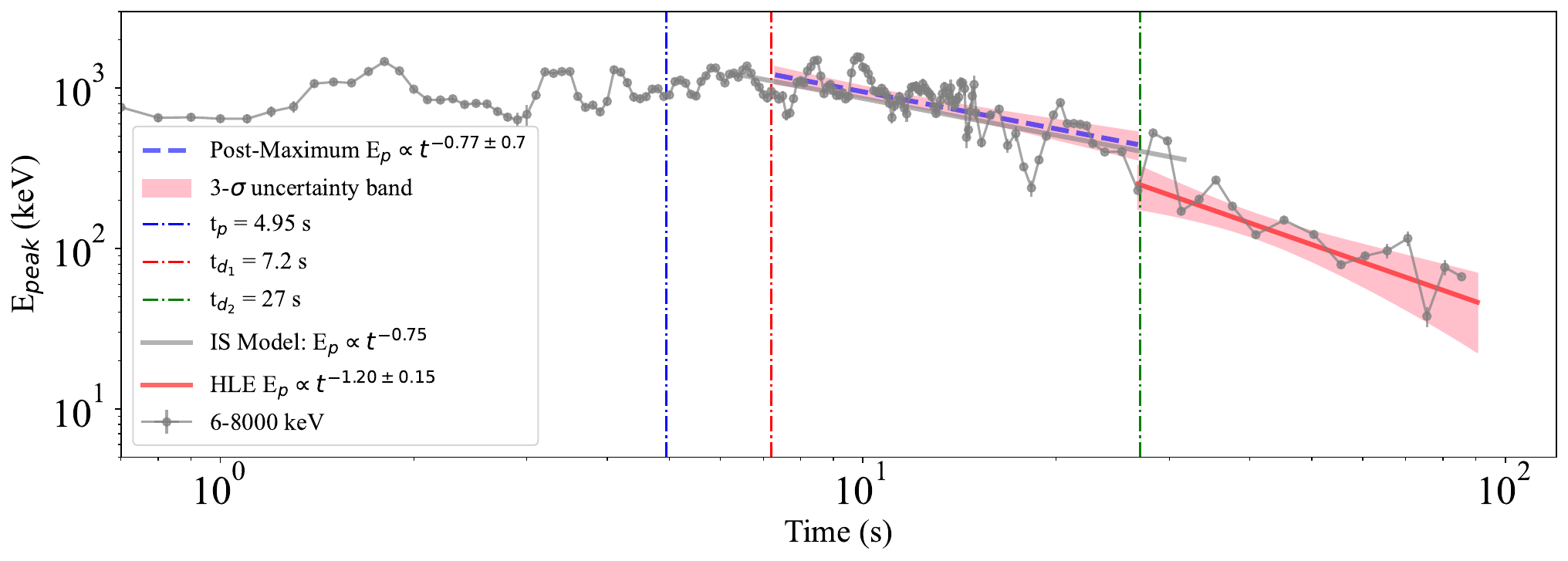} 
\includegraphics[width=0.99\hsize,clip]{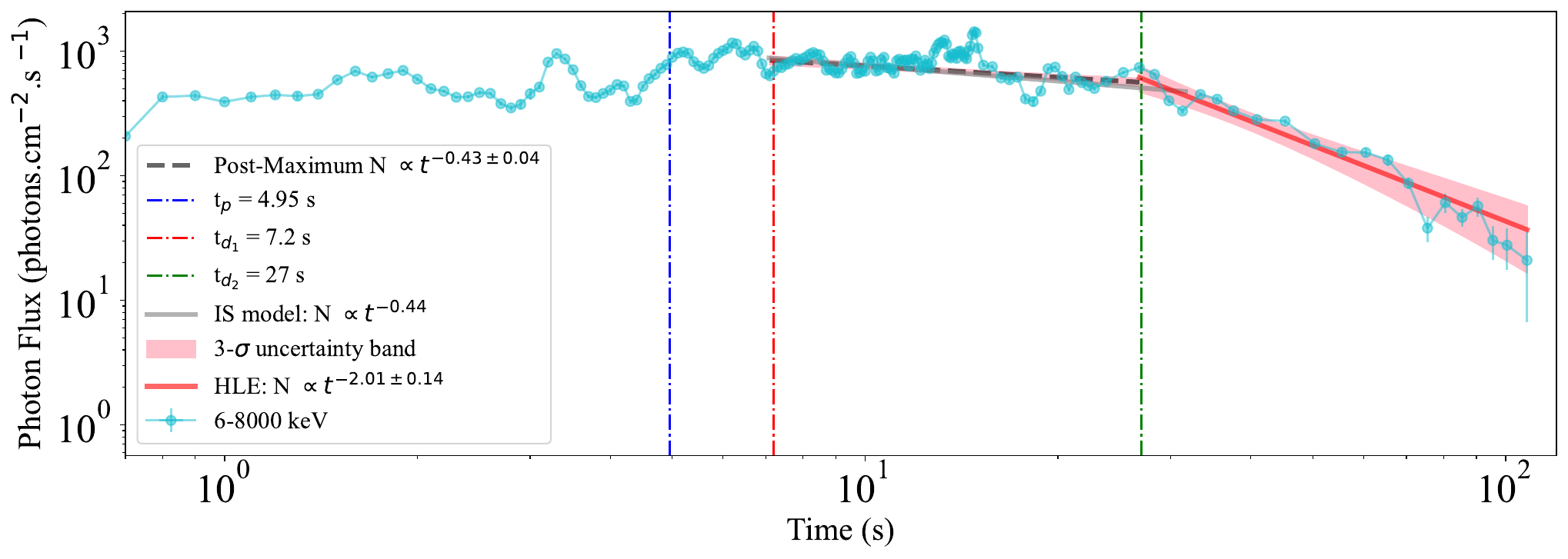} 

\caption{The best fit parameters of $F_{E}(t)$, $E_\mathrm{p}(t)$, and $N(t)$ during the decay phase of GRB 230307A are depicted in the plot. It is evident that fromt $t_{\rm d_1}$ = 7.2 to $t_{\rm d_2}$ = 27 s, the best fit results align closely with the predictions of the internal shock model. Beyond $t_{\rm d_2}$ = 27 s, the observed emission corresponds to the High-Latitude Emission (HLE) phase. }\label{fig:Model:Fits} 
\end{figure*}

Within the range of $t_{\rm d_1}$ = 7.2 s to $t_{\rm d_2}$ = 27 s, we observe that $F_{E}(t) \propto 1/t^{(1.55 \pm 0.10)}$, $E_\mathrm{p}(t) \propto  1/t^{(0.77\pm 0.07)}$, $N(t) \propto  1/t^{(0.43\pm 0.04)}$, and $\varphi_\mathrm{p} (E_ 
\mathrm{p})\propto E_\mathrm{p}^{(-0.49\pm 0.03)}$; see Figs.~\ref{fig:Model:N} and ~\ref{fig:Model:Fits}. 

These lead to HIC and HFC relations

\begin{equation}
E_\mathrm{p}(t)\propto N(t)^{(1.52\pm 0.03)}
\label{eq:HIC1}
\end{equation}
and 
\begin{equation}
E_\mathrm{p}(t)\propto e^{-\Phi_{N}(t)/(560\pm 2)}\ ,
\label{eq:HFC1}
\end{equation}
as shown in Fig.~\ref{fig:Model:N}. Moreover, the correlation between the peak energy $E_\mathrm{p}(t)$ and the associated energy flux, $F_{E}(t)$, believed to be intrinsic to the emission process, is consistent with the correlations identified by \textcolor{black}{\cite{2004ApJ...606L..29L}, \cite{2009A&A...496..585G}, \cite{2012ApJ...756..112L}, \cite{2012ApJ...754..138F}, and \cite{2018AdAst2018E...1D}}.

Beyond $t_{\rm d_2}$ = 27 s, the decay behavior follows $F_{E}(t) \propto  1/t^3$, $N(t) \propto  1/t^2$ and $E_\mathrm{p}(t) \propto 1/t^{1.1}$; see Fig.~\ref{fig:Model:Fits}. This temporal behaviour aligns with the characteristics of high latitude emission (HLE), which persists after the on-axis emission from the final dissipating regions in the relativistic outflow ceases; \textcolor{black}{\citep[see also][]{2000ApJ...541L..51K,2004ApJ...614..284D, 2005MNRAS.362...59S,2009ApJ...690L..10Z, 2013ApJ...778....3S,2015ApJ...808...33U, 2023arXiv230705689S}}. Such a HLE emission is in general consistent with the cessation of synchrotron emission in a large emission radius, such as the ICMART model \citep{2011ApJ...726...90Z,2021ApJS..253...43L,2024ApJ...963L..30U} and the large-radius internal shock model \citep{2012A&A...542L..29H} (but see \cite{2017SSRv..212..409D}). The photosphere models \citep{2008ApJ...682..463P,2011ApJ...737...68B} have a much smaller emission radius, so that the high-latitude emission timescale is expected to be much shorter and exhibit a rapid decay \citep{2014ApJ...785..112D}, which is inconsistent with the data. \textcolor{black}{Furthermore, the spectral analysis presented in Table \ref{tab:fit_result} shows no evidence of statistically significant thermal components, which are essential indicators of photospheric emission in such analyses.}

In the subsequent analysis, it is therefore crucial to investigate the decay behavior of $F_\mathrm{E}(t)$, $E_\mathrm{p}(t)$, and $N(t)$ within a general framework of synchrotron radiation at a large emission radius. A practical model is the large radius internal shock model \citep{2012A&A...542L..29H,2014A&A...568A..45B}, because it has less parameters than the ICMART model that invokes multiple minijets (see \cite{2014ApJ...782...92Z} and \cite{2022ApJ...927..173S} for detailed modeling) and because the two models share similar global properties in terms of $F_\mathrm{E}(t)$, $E_\mathrm{p}(t)$, and $N(t)$ evolutions. We want to, however, emphasize the caveats of the IS model. Since additional rapid variability is observed on top of the broad pulse profile of GRB 230307A, in order to fully interpret the data within the IS model, additional small-radius shocks are needed to account for the rapid variability. The superposition between small-radius ISs and a large-radius IS is actually disfavored by the data of GRB 230307A \citep{2023arXiv231007205Y}. Nonetheless, the goal of the following investigation is to see how good the simplest IS model can reproduce the general behavior of this burst.

\begin{figure}
\centering
\includegraphics[width=1\hsize,clip]{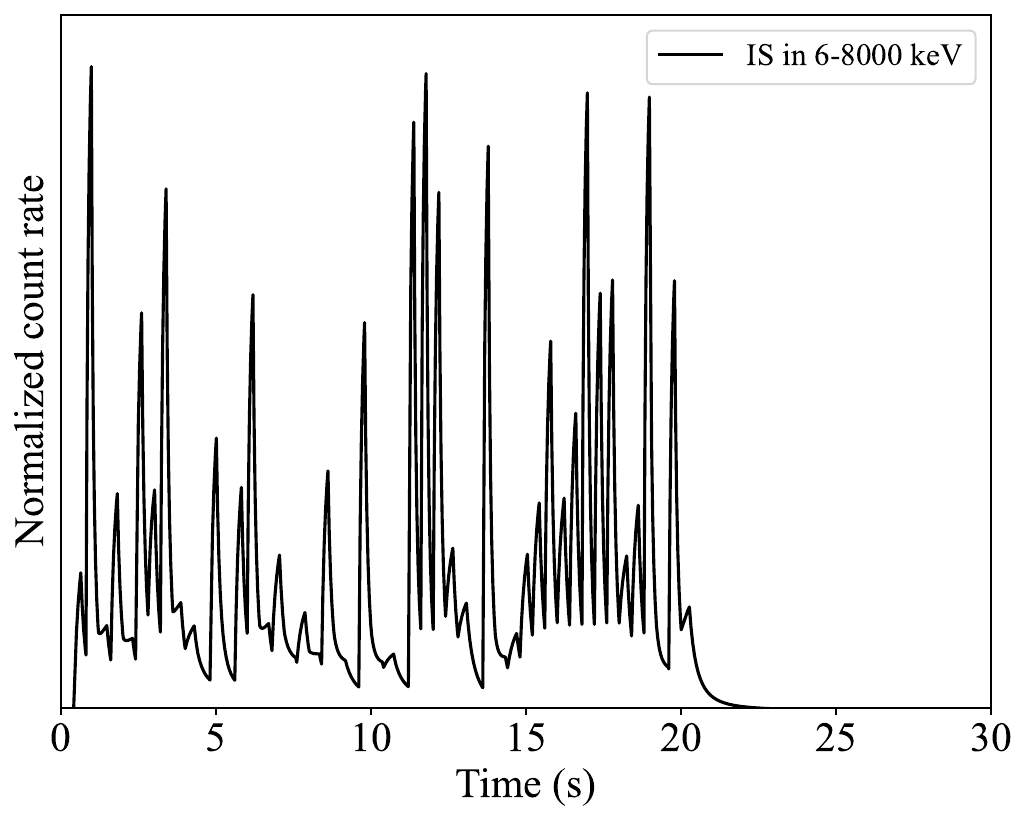} 
\caption{The plot represents the normalized count rate obtained from the simulation of randomized shell collisions in the IS model versus time. 50 shells are ejected, each characterized by randomly chosen initial thickness, Lorentz factor, and mass from log-distributed random distributions. The Lorentz factors range from 100 to 1000, the mass spans from $10^{28}$ to $10^{29}$, and the initial thickness varies from $10^{10}$ to $2 \times 10^{10}$, all in cgs units. The ejection times from the central engine follow a linear distribution, with values ranging from 0 to 20 seconds in the rest frame of the GRB central engine. }\label{fig:Model:IS-Random} 
\end{figure}

%%%%%%%%%%%%%%%%%%%%%%%%%%%%%%%%%%%%%%%%%%%%%%%%%%%%
%%%%%%%%%%%%%%%%%%%%%%%%%%%%%%%%%%%%%%%%%%%%%%%%%%%%
\section{Internal Shock Model Description}\label{sec:3}
%%%%%%%%%%%%%%%%%%%%%%%%%%%%%%%%%%%%%%%%%%%%%%%%%%%%
%%%%%%%%%%%%%%%%%%%%%%%%%%%%%%%%%%%%%%%%%%%%%%%%%%%%
The internal shock model, introduced by \cite{1994ApJ...430L..93R}, explains the GRBs originating from compact stellar-mass objects. It suggests that a wind emitted by such objects has an uneven distribution of speed, resulting in faster sections catching up with slower ones, generating gamma-rays through radiative processes such as synchrotron or inverse Compton radiation through internal shocks \citep{1996AIPC..384..782S,1997MNRAS.287..110S,1997ApJ...490...92K, 1998MNRAS.296..275D}; \cite[see also][for more information about development of the internal shock model.]{zhang2018physics, 2022Galax..10...38B,2009A&A...498..677B,2014A&A...568A..45B}. 

%%%%%%%%%%%%%%%%%%%%%%%%%%%%%%%%%%%%%%%%%%%%%%%%%%%%
%%%%%%%%%%%%%%%%%%%%%%%%%%%%%%%%%%%%%%%%%%%%%%%%%%%%
\subsection{Discrete randomize shell collisions}\label{sec:3.1}
%%%%%%%%%%%%%%%%%%%%%%%%%%%%%%%%%%%%%%%%%%%%%%%%%%%%
%%%%%%%%%%%%%%%%%%%%%%%%%%%%%%%%%%%%%%%%%%%%%%%%%%%%

 Internal shocks (ISs) were initially thought to be discrete shell collisions \citep{1997ApJ...490...92K}, but this led to discrepancies with observed GRB decay rates due to the ``curvature effect.'' In the multiple shells IS scenario, the long-term overall trend of the light curve reflects the activity history of the central engine, showing neither shape-energy dependence nor any power-law behavior in the luminosity trend. This random behavior is not surprising, as the cluster of shells is randomly generated and possesses random mass and velocity.

 Within a relativistic wind characterized by a nonuniform Lorentz factor, internal shocks occur and transform a fraction of the kinetic energy into radiation. Our approach involves depicting the irregular wind as a series of relativistic shells, with the collision of two shells constituting the fundamental process in this model. When multiple shells are present, numerous collisions occur. To delineate the temporal structure, we compute the time sequence of these two-shell collisions and overlay the resultant pulses from each collision.

A rapidly-moving shell (denoted by the subscript r) overtakes a slower one (s), and the two combine to create a single merged shell (m). This system mimics an inelastic collision between two masses $m_{\rm r}$ and $m_{\rm s}$. By applying the principles of energy and momentum conservation, we determine the Lorentz factor of the merged shell to be \citep{1997ApJ...490...92K}:
\begin{equation}
\gamma_{\rm m} \simeq \sqrt{\frac{\gamma_{\rm r} m_{\rm r}+ \gamma_{s} m_{\rm s}}
{\frac{m_{\rm r}}{\gamma_{\rm r}}+\frac{m_{\rm s}}{\gamma_{\rm s}}}}.
\label{momentum}
\end{equation}

Subsequently, each collision releases an internal energy as described by: 

\begin{equation}
E_{\rm int}=(\gamma_{\rm r} - \gamma_{\rm m})m_{\rm r} c^{2} + (\gamma_{\rm s} -
\gamma_{\rm m}) m_{\rm s} c^{2}.
\end{equation}
The efficiency of each collision can be defined as
\begin{equation}
\eta = \frac{E_{\rm int}}{(\gamma_{\rm s} m_{\rm s} c^{2} + \gamma_{\rm r} m_{\rm r} c^{2})}~.
\end{equation}

The luminosity observed for each shell, denoted as $\mathcal{L}(t)$, is expressed as:
\begin{equation}
\mathcal{L}(t)  = \left\{
\begin{array}{l l}
  h[1-1/(1+2\gamma^2_{\rm m} c t/R)^2], & \quad 0 < t <\delta t_e/2\gamma^2_{\rm m} \\
  \\
  h(1/[1+(2\gamma^2_{\rm m} t - \delta t_{\rm e})c/R] \\ - 1/(1+2\gamma^2_{\rm m} c t/R)^2), &\quad t > \delta t_e/2\gamma^2_{\rm m} .\\ \end{array} \right.\
\label{shellspreading}
\end{equation}
where $h = E_{\rm int} 2\gamma^2_{\rm m}/ \delta t_{\rm e}$, $R$ represents the collision radius, and $\delta t_e$ denotes the emission timescale at the moment when the reverse shock crosses the fast-moving shell. This is expressed as $\delta t_{\rm e} = l_{\rm r}/c(\beta_{\rm r} - \beta_{\rm rs})$, where $l_{\rm r}$ is the width of the rapid shell, $\beta_{\rm r}=(1-1/\gamma_{\rm r}^2)^{1/2}$, and  $\beta_{rs}=(1-1/\gamma_{\rm rs}^2)^{1/2}$, being  $\gamma_{\rm rs}$ the Lorentz factor of reverse shock approximated as:
\begin{equation}
\gamma_{\rm rs} \simeq \gamma_{\rm m} \sqrt{(1+\frac{2\gamma_{\rm m} }{\gamma_{\rm r}})/(2+\frac{\gamma_{\rm m}}{\gamma_{\rm r}})}.
\label{rs}
\end{equation}

We investigate the collisions of a cluster of shells released in a single emission event, specifically relevant to GRB 230307A, characterized by immediate emission without distinct intervals between pulses. By simulating the ejection of multiple randomized shells from the central engine, we are able to closely monitor the properties of each individual shell. For instance, we eject 50 shells, with their initial thickness, Lorentz factor, and mass chosen from random distributions in log space. The Lorentz factors vary from $100$ to $1000$, the mass ranges from $10^{28}$ to $10^{29}$, and the initial thickness ranges from $10^{10}$ to $2 \times 10^{10}$, all in cgs units. The central engine's ejection times are randomly selected from a linear distribution, with values ranging from $0$ to $20$ seconds in the rest frame of the GRB central engine. Upon collision, two shells merge, and the merged shell parameters ($m$, $\gamma$, $l_{\rm r}$) are adopted as the new values for the remaining shell. To ensure the tracking of future collisions, the ``effective" ejection time of this new shell is redefined as $t_{\rm ej,m}=R/c\beta_{\rm m}$, where $\beta_{\rm m}=(1-1/\gamma_{\rm m}^2)^{1/2}$. The simulation then restarts with one less shell. This process is repeated for each collision, allowing the code to monitor all collision/merging events for any arbitrarily designed central engine activity \citep{1997ApJ...490...92K,Maxham-Zhang09}.

To determine $E_\mathrm{p}$, we utilize an empirical relationship between the isotropic emission energy $E_{\rm iso}$ and $E_{\rm p}$, which has been found to be widely applicable to GRBs \citep{2002A&A...390...81A, 2009ApJ...704.1405K} and within the time-resolved analysis of individual bursts \citep{2004ApJ...606L..29L, 2009A&A...496..585G}. This relationship has also been demonstrated to be valid in GRB 230307A, as illustrated in Fig.~\ref{fig:Model:N}. We assume the validity of this correlation, represented by the equation:

\begin{equation}
E_{\rm p} =100 ~{\rm keV} \left( \frac{E_{\rm iso}}{10^{52}~{\rm erg}} \right) ^{1/2}   \ . 
\end{equation}

We then use $E_{\rm iso}=E_{\rm int}$ to estimate $E_{\rm p}$.

The photon flux $N(t)$ in the energy channel [$E_\mathrm{1,obs};E_\mathrm{2,obs}]$ is approximately given by Eq.~\ref{eq:FE}:

\begin{equation}
N_{12} (t)\simeq \frac{\mathcal{L}_F(t)}{4\pi D_\mathrm{L}^2 E_\mathrm{p}(t) \textcolor{black}{ \varphi_{0}}} \times \int_{E_\mathrm{1,obs}/E_\mathrm{p,obs}}^{E_\mathrm{2,obs}/E_\mathrm{p,obs}} \mathcal{N}(x) dx \ ,
\end{equation}
where $D_\mathrm{L}$ denotes the luminosity distance, $\mathcal{L}_F(t)$ represents the superposition of all $\mathcal{L}(t)$ from each collision as shown in Eq.~\ref{shellspreading}.

For GRB 230307A, the spectral shape is described by the CPL model, defined as $\mathcal{N}(E)=A(\frac{E}{E_{\rm piv}\,{}})^{\alpha}{\rm exp}(-\frac{E}{E_{\rm c}})$, where $A$ is the normalization factor in units of ph cm$^{-2}$keV$^{-1}$s$^{-1}$, $E_{\rm piv}$ is the pivot energy fixed at $100$~keV, $\alpha$ is the low-energy power-law photon spectral index, and $E_{c}$ is the break energy in units of keV listed in Table.~\ref{tab:fit_result}. The simulation results for [$ E_\mathrm{1,obs}$ = $\rm 6~keV$;$E_\mathrm{2,obs}$ = $\rm 8000~keV]$ are depicted in Fig.~\ref{fig:Model:IS-Random}. The light curve does not follow a specific trend, and the trend changes for each set of random variables.

%%%%%%%%%%%%%%%%%%%%%%%%%%%%%%%%%%%%%%%%%%%%%%%%%%%%
%%%%%%%%%%%%%%%%%%%%%%%%%%%%%%%%%%%%%%%%%%%%%%%%%%%%
\subsection{Continuous relativistic ejections}\label{sec:3.2}
%%%%%%%%%%%%%%%%%%%%%%%%%%%%%%%%%%%%%%%%%%%%%%%%%%%%
%%%%%%%%%%%%%%%%%%%%%%%%%%%%%%%%%%%%%%%%%%%%%%%%%%%%

As outlined in the preceding subsection regarding the multiple shells IS scenario, where a cluster of shells is generated randomly with varying mass and velocity, the extended light curve demonstrates neither a correlation between shape and energy nor any power-law pattern in luminosity trends. To address this, a different model proposed by \cite{1998MNRAS.296..275D, 2003MNRAS.342..587D} considers continuous relativistic ejections, resulting in shock waves propagating within the outflow. This modified approach effectively considers one single collision at a large emission radius, which better matches the observed data, showing slower pulse decays \citep{2014A&A...568A..45B}. The simplified model in this context by \cite{1998MNRAS.296..275D, 2003MNRAS.342..587D} approximates the flow evolution, treating the wind as solid shells interacting through collisions, while disregarding pressure waves present in a full hydrodynamical scenario \citep{2000A&A...358.1157D}.

In our investigation of the decaying phase of the burst, we employ a simplified pulse analytic model proposed by \cite{2003MNRAS.342..587D}. Given that our focus is solely on the overall decaying behavior and not on calculating variability in this particular burst \citep{2023ApJ...954L..29D}, we have chosen to utilize this analytical model in this work due to its advantageous analytical properties.

The analytical solution involves a relativistic wind, wherein at time $ t = t_{0}$ a slower shell with mass $M_0$ and Lorentz factor $\Gamma_0$ starts decelerating a faster portion of the flow characterized by a constant rest-frame mass flux $\dot{M}$ and Lorentz factor $\Gamma_1>\Gamma_0$. Being $t$ the observer time, and $\gamma=\Gamma/\Gamma_1$, with $\Gamma$ and $M$ being the current Lorentz factor and mass of the slow shell. Due to the accretion of rapidly moving material, the Lorentz factor of the slow shell increases. When a mass element $\mathrm{d}M$ is accreted, the Lorentz factor becomes

\begin{equation}
\Gamma+\mathrm{d}\Gamma=\left(\Gamma_1\Gamma {\Gamma_1 \mathrm{d}M+\Gamma M\over
\Gamma\mathrm{d}M+\Gamma_1 M}\right)^{1/2}\ ,
\end{equation}
and the dissipated power can be approximated by 
\begin{eqnarray}
     & \dot{\mathcal{E}}(\tau) = \frac{dM}{dt} \Gamma\epsilon c^2 = \frac{\dot{M}\Gamma_1 c^2}{2} 
(1-\gamma^2)(1-\gamma)^2\ \nonumber \\ 
     & \overset{}{\xlongrightarrow{\rm at~large~\tau}}\propto \tau^{-1.5},
     \label{eq:power}
\end{eqnarray}
where $\epsilon c^2$ the dissipated energy per unit mass in the comoving frame, $t_0=M_0/\dot{M}$, $\tau=t/t_0$, and initial $\gamma_0=\Gamma_0/\Gamma_1$ \citep[see][]{2003MNRAS.342..587D}. For decaying part, $\gamma(\tau)$ is approximated by
\begin{equation}
\gamma(\tau) \simeq 1-\frac{1}{2\sqrt{Q\tau}}\ ,
\label{eq:approx}
\end{equation}
with $Q=\frac{1}{2}\left(\frac{1+\gamma_0}{1-\gamma_0}\right)$. 
%

%%%%%%%%%%%%%%%%%%%%%%%%%%%%%%%%%%%%%%%%%%%%%%
\section{Emission Processes}\label{sec:4}
%%%%%%%%%%%%%%%%%%%%%%%%%%%%%%%%%%%%%%%%%%%%%%

Now, we examine the spectral evolution during the pulse decay and explore the underlying emission processes. Following \cite{1998MNRAS.296..275D, 2003MNRAS.342..587D} we consider that the dissipated energy in the system is radiated solely through the synchrotron process. The peak energy ($E_\mathrm{p}$) is influenced by the Lorentz factor ($\Gamma$), magnetic field strength ($B$), and the characteristic electron Lorentz factor behind the shock ($\Gamma_\mathrm{e}$). Classical equipartition assumptions allow to express the magnetic field and the characteristic electron Lorentz factor as functions of the comoving density ($\rho$) and the dissipated energy per unit mass ($\epsilon c^2$). Specifically, for the standard synchrotron radiation within internal shock model the peak energy can be written as 
\begin{equation}
E_\mathrm{p} = E_\mathrm{syn} \propto \Gamma~B ~\Gamma^2_\mathrm{e}  \propto \Gamma\rho^{1/2} \epsilon^{5/2} \propto \frac{(1-\gamma)^5}{\gamma^{7/2}t}\ ,
\end{equation}
which in the late decaying phase behaves as a power law: $E_\mathrm{p}\propto t^{-7/2}$. However, empirical observations obtained from the best fit of the decay phase of GRB 230307A, as shown in section~\ref{sec:2}, show that the observed spectral evolution of pulses deviates from this steep power law behavior at late times and decays as $E_\mathrm{p}\propto t^{-0.77}$.

To account for the observations, a more generalized phenomenological expression for the peak energy is introduced, given by $E_\mathrm{p} \propto \Gamma \rho^x \epsilon^y  \propto \frac{(1-\gamma)^{2y}}{\gamma^{4x+y-1}t^{2x}}\ $ which \textit{at later time} becomes
\begin{equation}\label{eq:epn}
 E_\mathrm{p} \propto \frac{1}{t^{2x+y}}\ ,  
\end{equation}
where $x$ and $y$ are exponents that can differ from the standard synchrotron values of $1/2$ and $5/2$, \citep{1996ApJ...461L..37B,2005ApJ...627..861E,2008ApJ...682L...5S}.

The values of $x$ and $y$ are crucial in characterizing the observed spectral evolution of GRBs. For the case of GRB 230307A, when $x=y=1/4$, the observed peak energy, $E_\mathrm{p} \propto 1/t^{0.77}$ is reproduced. In the following subsection, we will investigate the significance of these values for $x$ and $y$. 

%%%%%%%%%%%%%%%%%%%%%%%%%%%%%%%%%%%%
%%%%%%%%%%%%%%%%%%%%%%%%%%%%%%%%%%%%
\subsection{Temporal profiles of Flux and photon flux}\label{sec:4.1}
%%%%%%%%%%%%%%%%%%%%%%%%%%%%%%%%%%%%
%%%%%%%%%%%%%%%%%%%%%%%%%%%%%%%%%%%%
 We have so far investigated the behavior of spectral evolution during pulse decay and presented a more versatile expression for the peak energy, considering different exponents to match empirical observations. Now, we are going to show that this choice of parameters is also consistent with the observed flux, and photon flux of GRB 230307A, when considering the internal shock model. The synthetic pulse profile depicted in Fig.~\ref{fig:Model:Fits} is generated by decelerating a wind with fast shells with Lorentz factor $\Gamma_1=400$ and total power $\dot{M}\Gamma_1 c^2=10^{52}$ erg.s$^{-1}$ decelerated by slow shells with Lorentz factor $\Gamma_0=100$ as illustrated in section \ref{sec:3}. The adopted values are $t_0=0.6$ s, $x=y=1/4$. The pulse duration aligns with the expectation of being approximately $27$~s.

The post-maximum evolution of the flux at 6-8000 keV bandwidth exhibits a decay behavior of $1/t^{1.5}$, as depicted in Fig.~\ref{fig:Model:Fits}. This decay pattern can be comprehended by employing Eq. \ref{eq:power}, when combined with the Lorentz gamma profile of Eq. \ref{eq:approx}. The behavior of $N(t) \propto 1/t^{0.44}$ within internal shock model is calculated using Eq. \ref{eq:FE} in conjunction with Eq. \ref{eq:power}, which aligns with the observation, $N(t) \propto 1/t^{0.43}$, as shown in Fig.~\ref{fig:Model:Fits}.

From our analysis, we have determined that the discrepancies observed in the spectral evolution of GRB 230307A compared to the predictions of the internal shock model proposed in this study may stem from oversimplified assumptions regarding the microphysics within the emitting shocked regions. Additionally, these discrepancies could be attributed to the nature of the magnetic field, which may naturally decay with radius or be influenced by the presence of an external magnetic field \cite{2016ApJ...825...97U,2023ApJ...957..109Z}.

\subsection{More complex observation and the internal-collision-induced magnetic reconnection and turbulence (ICMART) model}

The current understanding of relativistic shocks does not offer a detailed description. However, incorporating fluctuations in the fraction of accelerated electrons based on shock intensity allows for achieving more qualitative and quantitative agreement between model predictions and observed spectral evolution for many GRBs \citep[see][for a comprehensive investigation of parameter selection concerning these discrepancies]{2014A&A...568A..45B}.

The issues raised above may be more complex than initially thought. Through the use of time-resolved spectral analysis of GECAM data, \citet{2023arXiv231007205Y} have demonstrated that the broad pulse of GRB 230307A consists of numerous rapidly variable short pulses, rather than being caused by short pulses overlaid on top of a slow component. This poses a significant challenge to the internal shock models, which attribute all variability components to collisions among different shells. Instead, they propose the ICMART model \citep{2011ApJ...726...90Z}, wherein the prompt radiation is generated by many mini-jets resulting from local magnetic reconnection events in a large emission zone far from the GRB central engine. Most notably, GRB 230307A shows the presence of a dip at t $\sim$ 18~s in all energy bands observed by GECAM and Fermi/GBM. During the decay phase, significant dip with its location consistent with such a decay phase, is allowed in the ICMART model. Because the fast pulses originate from local mini-jet events, the broad-band emission of the fast component is related to the dynamics of the mini-jets and is therefore aligned in time. 

The presence of a dip and numerous rapidly variable short pulses in the prompt emission of GRB 230307A challenges the traditional approach to understanding the pulse behavior of these new class of GRBs and calls for a revision. It also emphasizes the need to consider models other than the internal shock model to explain such peculiar behavior. Further observations of similar GRBs with high-resolution detectors, such as GECAM, are necessary to draw a general conclusion.

%%%%%%%%%%%%%%%%%%%%%%%%%%%%%%%%%%%%%%%%%%%%%%%%%%%%
%%%%%%%%%%%%%%%%%%%%%%%%%%%%%%%%%%%%%%%%%%%%%%%%%%%%
\section{Conclusions}\label{sec:5}
%%%%%%%%%%%%%%%%%%%%%%%%%%%%%%%%%%%%%%%%%%%%%%%%%%%%
%%%%%%%%%%%%%%%%%%%%%%%%%%%%%%%%%%%%%%%%%%%%%%%%%%%%

In this paper, we presented the pulse profile of GRB 230307A and conducted a detailed analysis of its temporal behavior using a combined GECAM--\textit{Fermi}/GBM time-resolved spectral analysis. The inclusion of GECAM data is particularly advantageous as it effectively captured crucial information during the pile-up period of the \textit{Fermi}/GBM \citep{2023GCN.33551....1D}, showcasing its unique ability to gather data from highly luminous GRBs. This capability is especially valuable in enhancing our understanding of highly luminous GRBs, such as GRB 230307A and GRB 221009A \citep{2023ApJ...952L..42L}.

The analysis performed in this study is essential in validating the existence of the observed hardness-intensity correlation (HIC) and hardness-flux correlation (HFC). These correlations are pivotal observational relations in GRBs, as they can aid in distinguishing among different theoretical models.

The proposed multiple shells IS scenario, as outlined by \cite{1997ApJ...490...92K}, displays a light curve trend that mirrors the historical activity of the central engine. This trend does not exhibit a dependence on shape-energy or adhere to a power-law behavior in luminosity.  This stochastic behavior is expected, considering the random generation of the shell cluster, which possesses varied mass and velocity parameters. Achieving precise alignment of all elements in this model is challenging due to its inherent randomness, necessitating a highly specialized design to ensure accurate representation of the system.

We have observed that the decay phase  of GRB 230307A is consistent with the large-radius synchrotron emission model, in particular, within the context of the large-radius internal shock model \citep{1998MNRAS.296..275D, 2003MNRAS.342..587D}, which posits continuous relativistic ejections causing shock waves to propagate within the outflow. However, the parameter values $x = 1/4$ and $y = 1/4$ are lower than the expected values $x = 1/2$ and $y = 5/2$ for standard synchrotron radiation.

This deviation suggests in the internal shock framework a potential scenario where the correlation between the magnetic field, outflow Lorentz factor, electron Lorentz factor, and the dissipated energy is less pronounced. As highlighted by \citet{2014A&A...568A..45B}, uncertainties persist in understanding the microphysics involved in shocked material, which could be attributed to oversimplified assumptions about the microphysics within the emitting shocked regions \citep{2014A&A...568A..45B}.
On the other hand, this choice of parameters is consistent with a more general model proposed by \cite{2016ApJ...825...97U}, which involves a Poynting-flux-dominated jet dissipating magnetic energy at a distance from the engine. In this model, the magnetic field strength decreases with radius as the emission region expands, and rapid bulk acceleration occurs during the production of prompt $\gamma$-rays.
Alternatively, it indicates the possible need to incorporate additional elements, such as a combination of a decaying shock-generated magnetic field and a non-decaying background magnetic field, to provide a better fit to the observations compared to the standard internal shocks model \citep{2023ApJ...957..109Z}.

On the other hand, \citet{2023arXiv231007205Y} demonstrated that the broad pulse of GRB 230307A is actually composed of numerous rapidly variable short pulses, rather than being the result of the superposition of many short pulses on top of a slow component. This finding aligns with the concept of many mini-jets arising from local magnetic reconnection events in a large emission zone far from the GRB central engine, as proposed in the ICMART model. This challenges the internal shock models, which attribute all variability components to collisions among different shells. Consequently, it is not surprising that the internal shock model for this GRB requires ad hoc assumptions in order to provide an acceptable pulse explanation.

An area requiring further investigation in the internal shock model is the reliability of the initial Lorentz factor profile. For example, it is becoming increasingly accepted that the kinetic energy of ejecta in GRBs is powered by the rotational energy of the central engine. However, the utilization of rotational energy-powered ejecta results in a reduction of the overall rotational energy of the central engine, consequently leading to a smaller initial kinetic energy for the remaining ejecta. As a consequence, this affects the temporal decrease of the Lorentz gamma factor of the ejecta \citep[see e.g.][for more details on the rotational energy-powered ejecta in GRBs]{2021PhRvD.104f3043M, 2022EPJC...82..778R}, raising concerns about the validity of the initial profile employed in the internal shock model, which highly relies on the deceleration of rapidly moving materials by slower materials within a relativistic wind. 

\section*{Acknowledgements}
\textcolor{black}{We would like to thank the anonymous referee for his/her insightful comments, which have significantly improved the presentation of the paper and the quality of the analysis performed.} R. Moradi acknowledges support from the Institute of High Energy Physics, Chinese Academy of Sciences (E32984U810).
This work is supported by the Beijing Natural Science Foundation (IS24021).



\bibliographystyle{aasjournal}

\clearpage
\begingroup\tiny
\begin{ThreePartTable}
\begin{TableNotes} 
\item[*] 
The PL model is defined as  $N(E)=A(\frac{E}{100\rm{keV}})^{\alpha}$, \\
The CPL model is defined as  $N(E)=A(\frac{E}{100\,{\rm keV}})^{\alpha}{\rm exp}(-\frac{E}{E_{\rm c}})$, \\
The BAND model is defined as $N(E)=\left\{
 \begin{array}{l}
 AE^{\alpha_1}{\rm exp}(-\frac{E}{E_1}), ~E \leq E_{\rm b}, \\
 AE_{\rm b}^{\alpha_1-\alpha_2}{\rm exp}(\alpha_2-\alpha_1)E^{\alpha_2}{\rm exp}(-\frac{E}{E_2}), ~E > E_{\rm b}, \\
 \end{array}\right.$
\end{TableNotes} 
\begin{longtable}{@{}ccccccccc@{}}
\toprule
Time Interval & best model & $\alpha$ & $\beta$ & $E_{\rm p}$ & flux & photon flux & \textcolor{black}{CSTAT} & \textcolor{black}{DOF} \\ 
(s) & & & & (keV) & (erg$\cdot$s$\cdot$keV$\cdot$cm$^2$) & (photon$\cdot$s$\cdot$keV$\cdot$cm$^2$) && \\ 
\hline
\endhead 
\multicolumn{5}{r}{\textit{continued}}
\endfoot
\insertTableNotes \\ 
\endlastfoot 
-0.05-0.05 & band & -0.34$^{+0.08}_{-0.08}$ & -2.53$^{+0.07}_{-0.09}$ & 148.68$^{+5.77}_{-6.53}$ & 3.66$^{+2.00}_{-2.03}\times$10$^{-5}$ & 255.4$^{+15.39}_{-13.7}$ & 600 & 600 \\  
0.05-0.15 & band & -0.3$^{+0.05}_{-0.06}$ & -2.77$^{+0.07}_{-0.07}$ & 150.9$^{+3.96}_{-5.03}$ & 6.46$^{+2.20}_{-2.03}\times$10$^{-5}$ & 487.97$^{+21.66}_{-16.97}$ & 745 & 600 \\  
0.15-0.25 & band & -0.44$^{+0.06}_{-0.07}$ & -2.61$^{+0.07}_{-0.07}$ & 128.47$^{+4.16}_{-4.85}$ & 4.04$^{+1.88}_{-1.71}\times$10$^{-5}$ & 367.16$^{+20.34}_{-19.04}$ & 619 & 600 \\  
0.25-0.35 & cpl & -1.22$^{+0.08}_{-0.09}$ & - & 73.42$^{+3.57}_{-3.51}$ & 1.05$^{+8.02}_{-6.80}\times$10$^{-5}$ & 437.94$^{+73.41}_{-62.68}$ & 584 & 601 \\  
0.35-0.45 & pl & -2.08$^{+0.04}_{-0.04}$ & - & - & 1.42$^{+1.12}_{-1.08}\times$10$^{-5}$ & 1327.54$^{+232.07}_{-181.73}$ & 467 & 602 \\  
0.45-0.55 & pl & -1.93$^{+0.03}_{-0.03}$ & - & - & 1.44$^{+1.21}_{-1.03}\times$10$^{-5}$ & 805.95$^{+116.26}_{-102.28}$ & 602 & 602 \\  
0.55-0.65 & pl & -1.89$^{+0.03}_{-0.04}$ & - & - & 1.45$^{+1.20}_{-1.12}\times$10$^{-5}$ & 683.73$^{+112.67}_{-91.32}$ & 496 & 602 \\  
0.65-0.75 & cpl & -0.93$^{+0.03}_{-0.03}$ & - & 759.96$^{+42.94}_{-42.59}$ & 4.72$^{+2.63}_{-2.49}\times$10$^{-5}$ & 207.31$^{+12.51}_{-10.79}$ & 586 & 563 \\  
0.75-0.85 & cpl & -0.78$^{+0.04}_{-0.04}$ & - & 653.08$^{+39.87}_{-36.28}$ & 1.11$^{+5.54}_{-5.44}\times$10$^{-4}$ & 428.76$^{+22.36}_{-22.14}$ & 433 & 563 \\  
0.85-0.95 & cpl & -0.62$^{+0.05}_{-0.05}$ & - & 658.1$^{+37.17}_{-30.83}$ & 1.43$^{+6.66}_{-6.05}\times$10$^{-4}$ & 439.14$^{+22.03}_{-18.22}$ & 489 & 563 \\  
0.95-1.05 & cpl & -0.58$^{+0.04}_{-0.04}$ & - & 644.73$^{+30.93}_{-29.65}$ & 1.30$^{+5.80}_{-5.93}\times$10$^{-4}$ & 391.31$^{+15.73}_{-14.33}$ & 474 & 563 \\  
1.05-1.15 & cpl & -0.76$^{+0.04}_{-0.06}$ & - & 643.58$^{+44.54}_{-44.89}$ & 1.12$^{+5.62}_{-6.27}\times$10$^{-4}$ & 428.15$^{+27.15}_{-20.92}$ & 520 & 563 \\  
1.15-1.25 & cpl & -0.9$^{+0.05}_{-0.04}$ & - & 714.63$^{+56.32}_{-51.2}$ & 1.02$^{+6.20}_{-5.42}\times$10$^{-4}$ & 445.02$^{+26.54}_{-25.74}$ & 504 & 563 \\  
1.25-1.35 & cpl & -0.89$^{+0.05}_{-0.05}$ & - & 765.27$^{+63.96}_{-59.71}$ & 1.07$^{+6.44}_{-6.37}\times$10$^{-4}$ & 435.44$^{+31.24}_{-28.48}$ & 502 & 563 \\  
1.35-1.45 & cpl & -0.84$^{+0.04}_{-0.03}$ & - & 1066.07$^{+55.87}_{-52.43}$ & 1.67$^{+7.74}_{-7.95}\times$10$^{-4}$ & 450.7$^{+21.94}_{-20.04}$ & 514 & 563 \\  
1.45-1.55 & cpl & -0.54$^{+0.03}_{-0.03}$ & - & 1091.85$^{+40.16}_{-39.17}$ & 3.40$^{+1.31}_{-1.26}\times$10$^{-4}$ & 588.11$^{+15.38}_{-15.36}$ & 534 & 563 \\  
1.55-1.65 & cpl & -0.41$^{+0.03}_{-0.03}$ & - & 1077.53$^{+33.17}_{-31.02}$ & 4.53$^{+1.50}_{-1.56}\times$10$^{-4}$ & 690.53$^{+15.74}_{-14.73}$ & 498 & 563 \\  
1.65-1.75 & cpl & -0.44$^{+0.03}_{-0.03}$ & - & 1269.11$^{+41.54}_{-39.52}$ & 4.63$^{+1.57}_{-1.42}\times$10$^{-4}$ & 618.21$^{+14.14}_{-13.29}$ & 521 & 563 \\  
1.75-1.85 & cpl & -0.34$^{+0.03}_{-0.03}$ & - & 1463.77$^{+43.15}_{-39.56}$ & 6.22$^{+2.08}_{-1.93}\times$10$^{-4}$ & 659.06$^{+13.5}_{-14.21}$ & 538 & 563 \\  
1.85-1.95 & cpl & -0.27$^{+0.03}_{-0.03}$ & - & 1280.68$^{+36.79}_{-36.82}$ & 6.09$^{+1.96}_{-1.96}\times$10$^{-4}$ & 699.22$^{+13.92}_{-13.71}$ & 542 & 563 \\  
1.95-2.05 & cpl & -0.36$^{+0.03}_{-0.04}$ & - & 984.46$^{+33.6}_{-32.06}$ & 3.74$^{+1.37}_{-1.23}\times$10$^{-4}$ & 595.78$^{+14.6}_{-13.53}$ & 586 & 563 \\  
2.05-2.15 & cpl & -0.43$^{+0.04}_{-0.04}$ & - & 845.95$^{+34.9}_{-31.68}$ & 2.54$^{+1.01}_{-9.82}\times$10$^{-4}$ & 499.3$^{+14.7}_{-13.53}$ & 548 & 563 \\  
2.15-2.25 & cpl & -0.38$^{+0.03}_{-0.03}$ & - & 843.16$^{+28.11}_{-25.8}$ & 2.53$^{+9.23}_{-9.14}\times$10$^{-4}$ & 476.13$^{+11.2}_{-11.43}$ & 548 & 563 \\  
2.25-2.35 & cpl & -0.41$^{+0.04}_{-0.03}$ & - & 857.18$^{+35.61}_{-32.27}$ & 2.22$^{+9.73}_{-9.17}\times$10$^{-4}$ & 426.36$^{+11.91}_{-11.55}$ & 563 & 563 \\  
2.35-2.45 & cpl & -0.43$^{+0.04}_{-0.05}$ & - & 789.91$^{+31.58}_{-31.0}$ & 2.03$^{+8.30}_{-7.94}\times$10$^{-4}$ & 430.14$^{+12.38}_{-12.32}$ & 515 & 563 \\  
2.45-2.55 & cpl & -0.42$^{+0.04}_{-0.04}$ & - & 803.4$^{+29.13}_{-29.35}$ & 2.27$^{+8.71}_{-8.71}\times$10$^{-4}$ & 464.33$^{+13.97}_{-13.04}$ & 502 & 563 \\  
2.55-2.65 & cpl & -0.45$^{+0.05}_{-0.04}$ & - & 796.23$^{+34.11}_{-32.3}$ & 2.14$^{+8.85}_{-8.36}\times$10$^{-4}$ & 457.49$^{+12.16}_{-12.79}$ & 548 & 563 \\  
2.65-2.75 & cpl & -0.47$^{+0.05}_{-0.05}$ & - & 714.72$^{+36.57}_{-35.05}$ & 1.56$^{+7.49}_{-7.40}\times$10$^{-4}$ & 378.8$^{+14.54}_{-13.49}$ & 536 & 563 \\  
2.75-2.85 & cpl & -0.71$^{+0.04}_{-0.05}$ & - & 659.43$^{+40.03}_{-34.75}$ & 1.01$^{+5.62}_{-4.96}\times$10$^{-4}$ & 349.9$^{+19.57}_{-16.7}$ & 535 & 563 \\  
2.85-2.95 & cpl & -0.96$^{+0.05}_{-0.05}$ & - & 635.22$^{+74.77}_{-57.5}$ & 6.91$^{+5.24}_{-4.82}\times$10$^{-5}$ & 373.85$^{+31.98}_{-28.61}$ & 522 & 563 \\  
2.95-3.05 & cpl & -1.15$^{+0.05}_{-0.05}$ & - & 684.82$^{+116.54}_{-85.82}$ & 5.99$^{+5.57}_{-4.87}\times$10$^{-5}$ & 454.56$^{+50.5}_{-44.07}$ & 655 & 563 \\  
3.05-3.15 & cpl & -0.81$^{+0.04}_{-0.03}$ & - & 903.49$^{+54.38}_{-49.41}$ & 1.71$^{+8.14}_{-7.97}\times$10$^{-4}$ & 514.12$^{+22.48}_{-22.47}$ & 527 & 563 \\  
3.15-3.25 & cpl & -0.59$^{+0.03}_{-0.03}$ & - & 1262.23$^{+46.08}_{-41.71}$ & 5.12$^{+1.75}_{-1.69}\times$10$^{-4}$ & 815.97$^{+17.78}_{-19.19}$ & 535 & 563 \\  
3.25-3.35 & cpl & -0.44$^{+0.03}_{-0.03}$ & - & 1235.71$^{+34.85}_{-31.58}$ & 6.96$^{+1.99}_{-1.99}\times$10$^{-4}$ & 954.69$^{+19.09}_{-18.16}$ & 515 & 563 \\  
3.35-3.45 & cpl & -0.41$^{+0.03}_{-0.03}$ & - & 1268.14$^{+34.81}_{-31.69}$ & 6.64$^{+1.90}_{-1.94}\times$10$^{-4}$ & 863.85$^{+17.81}_{-17.0}$ & 507 & 563 \\  
3.45-3.55 & cpl & -0.46$^{+0.03}_{-0.03}$ & - & 1268.33$^{+42.42}_{-37.75}$ & 5.21$^{+1.67}_{-1.61}\times$10$^{-4}$ & 708.45$^{+14.93}_{-15.23}$ & 527 & 563 \\  
3.55-3.65 & cpl & -0.52$^{+0.03}_{-0.04}$ & - & 887.58$^{+38.66}_{-36.37}$ & 2.56$^{+1.06}_{-9.98}\times$10$^{-4}$ & 527.5$^{+16.05}_{-15.13}$ & 554 & 563 \\  
3.65-3.75 & cpl & -0.6$^{+0.04}_{-0.04}$ & - & 758.49$^{+34.95}_{-36.19}$ & 1.65$^{+7.62}_{-7.28}\times$10$^{-4}$ & 432.08$^{+17.09}_{-16.19}$ & 512 & 563 \\  
3.75-3.85 & cpl & -0.63$^{+0.04}_{-0.04}$ & - & 783.2$^{+38.41}_{-36.59}$ & 1.61$^{+7.51}_{-7.09}\times$10$^{-4}$ & 424.06$^{+15.12}_{-15.49}$ & 504 & 563 \\  
3.85-3.95 & cpl & -0.69$^{+0.04}_{-0.04}$ & - & 710.98$^{+36.52}_{-33.81}$ & 1.46$^{+6.15}_{-6.29}\times$10$^{-4}$ & 457.16$^{+19.38}_{-17.88}$ & 498 & 563 \\  
3.95-4.05 & cpl & -0.8$^{+0.04}_{-0.04}$ & - & 826.88$^{+51.11}_{-49.02}$ & 1.51$^{+7.77}_{-7.45}\times$10$^{-4}$ & 485.38$^{+23.68}_{-21.67}$ & 570 & 563 \\  
4.05-4.15 & cpl & -0.74$^{+0.03}_{-0.03}$ & - & 1300.57$^{+72.26}_{-61.67}$ & 2.84$^{+1.24}_{-1.23}\times$10$^{-4}$ & 539.13$^{+18.47}_{-18.17}$ & 544 & 563 \\  
4.15-4.25 & cpl & -0.61$^{+0.03}_{-0.03}$ & - & 1257.32$^{+60.37}_{-51.42}$ & 3.21$^{+1.26}_{-1.27}\times$10$^{-4}$ & 525.14$^{+15.96}_{-15.51}$ & 553 & 563 \\  
4.25-4.35 & cpl & -0.73$^{+0.04}_{-0.04}$ & - & 1080.83$^{+70.06}_{-62.9}$ & 1.77$^{+1.00}_{-8.87}\times$10$^{-4}$ & 394.64$^{+18.26}_{-17.35}$ & 477 & 563 \\  
4.35-4.45 & cpl & -0.84$^{+0.04}_{-0.04}$ & - & 880.9$^{+60.45}_{-52.65}$ & 1.25$^{+6.79}_{-6.39}\times$10$^{-4}$ & 404.82$^{+19.51}_{-19.26}$ & 444 & 563 \\  
4.45-4.55 & cpl & -0.75$^{+0.04}_{-0.04}$ & - & 858.41$^{+48.72}_{-43.76}$ & 1.83$^{+8.36}_{-8.23}\times$10$^{-4}$ & 521.1$^{+21.59}_{-20.08}$ & 459 & 563 \\  
4.55-4.65 & cpl & -0.75$^{+0.03}_{-0.03}$ & - & 884.48$^{+46.58}_{-42.42}$ & 2.15$^{+1.01}_{-8.69}\times$10$^{-4}$ & 600.59$^{+23.35}_{-21.08}$ & 541 & 563 \\  
4.65-4.75 & cpl & -0.67$^{+0.03}_{-0.04}$ & - & 985.13$^{+48.53}_{-41.76}$ & 2.91$^{+1.21}_{-1.12}\times$10$^{-4}$ & 650.37$^{+22.6}_{-18.99}$ & 580 & 563 \\  
4.75-4.85 & cpl & -0.56$^{+0.03}_{-0.03}$ & - & 991.06$^{+39.3}_{-34.39}$ & 3.77$^{+1.45}_{-1.37}\times$10$^{-4}$ & 727.34$^{+19.7}_{-18.25}$ & 591 & 563 \\  
4.85-4.95 & cpl & -0.65$^{+0.03}_{-0.03}$ & - & 887.15$^{+32.96}_{-30.25}$ & 3.25$^{+1.22}_{-1.12}\times$10$^{-4}$ & 779.21$^{+20.8}_{-21.46}$ & 590 & 563 \\ 
4.95-5.05 & cpl & -0.75$^{+0.03}_{-0.03}$ & - & 906.24$^{+37.83}_{-35.45}$ & 3.26$^{+1.17}_{-1.14}\times$10$^{-4}$ & 890.76$^{+28.79}_{-25.5}$ & 636 & 563 \\  
5.05-5.15 & cpl & -0.78$^{+0.02}_{-0.02}$ & - & 1098.62$^{+40.62}_{-41.23}$ & 4.02$^{+1.27}_{-1.37}\times$10$^{-4}$ & 962.55$^{+28.66}_{-27.1}$ & 553 & 563 \\  
5.15-5.25 & cpl & -0.71$^{+0.03}_{-0.02}$ & - & 1121.17$^{+39.58}_{-37.29}$ & 4.71$^{+1.52}_{-1.43}\times$10$^{-4}$ & 983.62$^{+24.87}_{-24.64}$ & 547 & 563 \\  
5.25-5.35 & cpl & -0.7$^{+0.03}_{-0.03}$ & - & 1072.72$^{+38.83}_{-39.42}$ & 4.43$^{+1.61}_{-1.44}\times$10$^{-4}$ & 959.01$^{+25.96}_{-26.06}$ & 575 & 563 \\  
5.35-5.45 & cpl & -0.71$^{+0.03}_{-0.03}$ & - & 914.78$^{+40.55}_{-34.73}$ & 3.22$^{+1.24}_{-1.07}\times$10$^{-4}$ & 821.76$^{+26.89}_{-24.57}$ & 538 & 563 \\  
5.45-5.55 & cpl & -0.77$^{+0.03}_{-0.03}$ & - & 892.64$^{+45.91}_{-39.66}$ & 2.65$^{+1.08}_{-1.01}\times$10$^{-4}$ & 760.5$^{+28.92}_{-26.74}$ & 540 & 563 \\  
5.55-5.65 & cpl & -0.78$^{+0.03}_{-0.03}$ & - & 1099.08$^{+46.8}_{-45.97}$ & 3.04$^{+1.10}_{-1.16}\times$10$^{-4}$ & 726.64$^{+26.59}_{-24.7}$ & 510 & 563 \\  
5.65-5.75 & cpl & -0.68$^{+0.02}_{-0.02}$ & - & 1193.34$^{+43.22}_{-42.05}$ & 4.06$^{+1.46}_{-1.43}\times$10$^{-4}$ & 765.57$^{+19.65}_{-20.09}$ & 605 & 563 \\  
5.75-5.85 & cpl & -0.61$^{+0.03}_{-0.02}$ & - & 1332.57$^{+43.23}_{-44.29}$ & 5.68$^{+1.76}_{-1.83}\times$10$^{-4}$ & 877.18$^{+20.98}_{-18.99}$ & 501 & 563 \\  
5.85-5.95 & cpl & -0.52$^{+0.02}_{-0.02}$ & - & 1335.13$^{+35.92}_{-35.89}$ & 6.95$^{+1.96}_{-1.86}\times$10$^{-4}$ & 958.52$^{+18.8}_{-19.24}$ & 525 & 563 \\  
5.95-6.05 & cpl & -0.57$^{+0.02}_{-0.02}$ & - & 1183.5$^{+35.51}_{-35.84}$ & 6.10$^{+1.79}_{-1.80}\times$10$^{-4}$ & 1010.42$^{+22.39}_{-20.6}$ & 545 & 563 \\  
6.05-6.15 & cpl & -0.62$^{+0.02}_{-0.02}$ & - & 1077.28$^{+33.93}_{-30.2}$ & 5.50$^{+1.63}_{-1.55}\times$10$^{-4}$ & 1049.47$^{+25.6}_{-22.19}$ & 483 & 563 \\  
6.15-6.25 & cpl & -0.61$^{+0.02}_{-0.02}$ & - & 1219.86$^{+34.25}_{-34.05}$ & 6.92$^{+1.85}_{-1.74}\times$10$^{-4}$ & 1163.55$^{+25.62}_{-24.25}$ & 583 & 563 \\  
6.25-6.35 & cpl & -0.59$^{+0.02}_{-0.02}$ & - & 1240.43$^{+35.32}_{-32.76}$ & 7.09$^{+2.03}_{-1.94}\times$10$^{-4}$ & 1141.44$^{+22.65}_{-23.59}$ & 583 & 563 \\  
6.35-6.45 & cpl & -0.58$^{+0.02}_{-0.02}$ & - & 1168.98$^{+37.1}_{-31.62}$ & 5.78$^{+1.69}_{-1.62}\times$10$^{-4}$ & 980.31$^{+19.65}_{-19.72}$ & 587 & 563 \\  
6.45-6.55 & cpl & -0.62$^{+0.02}_{-0.03}$ & - & 1297.27$^{+38.93}_{-40.45}$ & 5.79$^{+1.73}_{-1.72}\times$10$^{-4}$ & 930.43$^{+23.34}_{-21.62}$ & 599 & 563 \\  
6.55-6.65 & cpl & -0.57$^{+0.02}_{-0.02}$ & - & 1373.77$^{+44.22}_{-37.55}$ & 7.14$^{+2.16}_{-2.00}\times$10$^{-4}$ & 1014.46$^{+20.0}_{-21.42}$ & 543 & 563 \\  
6.65-6.75 & cpl & -0.58$^{+0.02}_{-0.02}$ & - & 1240.15$^{+37.15}_{-32.23}$ & 6.85$^{+2.05}_{-1.81}\times$10$^{-4}$ & 1092.46$^{+21.59}_{-21.5}$ & 542 & 563 \\  
6.75-6.85 & cpl & -0.64$^{+0.02}_{-0.03}$ & - & 1091.07$^{+36.0}_{-36.2}$ & 5.16$^{+1.63}_{-1.47}\times$10$^{-4}$ & 1001.36$^{+24.27}_{-22.99}$ & 579 & 563 \\  
6.85-6.95 & cpl & -0.7$^{+0.03}_{-0.03}$ & - & 1013.8$^{+43.87}_{-40.94}$ & 3.49$^{+1.33}_{-1.25}\times$10$^{-4}$ & 789.1$^{+27.73}_{-23.15}$ & 612 & 563 \\  
6.95-7.05 & cpl & -0.86$^{+0.03}_{-0.03}$ & - & 911.76$^{+51.81}_{-47.59}$ & 2.02$^{+9.03}_{-7.90}\times$10$^{-4}$ & 658.74$^{+31.22}_{-27.88}$ & 487 & 563 \\  
7.05-7.15 & cpl & -0.95$^{+0.04}_{-0.04}$ & - & 869.27$^{+59.68}_{-58.04}$ & 1.58$^{+7.52}_{-7.22}\times$10$^{-4}$ & 633.54$^{+37.69}_{-34.69}$ & 514 & 563 \\  
7.15-7.25 & cpl & -0.86$^{+0.03}_{-0.03}$ & - & 957.05$^{+58.25}_{-48.92}$ & 2.21$^{+9.96}_{-9.64}\times$10$^{-4}$ & 689.93$^{+33.38}_{-25.91}$ & 563 & 563 \\  
7.25-7.35 & cpl & -0.81$^{+0.03}_{-0.03}$ & - & 909.77$^{+42.43}_{-41.26}$ & 2.51$^{+9.68}_{-9.69}\times$10$^{-4}$ & 748.13$^{+30.72}_{-26.2}$ & 508 & 563 \\  
7.35-7.45 & cpl & -0.89$^{+0.03}_{-0.03}$ & - & 853.85$^{+45.01}_{-41.02}$ & 2.02$^{+8.45}_{-8.02}\times$10$^{-4}$ & 736.92$^{+34.89}_{-29.93}$ & 536 & 563 \\  
7.45-7.55 & cpl & -1.04$^{+0.03}_{-0.04}$ & - & 893.13$^{+78.4}_{-62.4}$ & 1.62$^{+8.62}_{-7.80}\times$10$^{-4}$ & 775.83$^{+50.88}_{-43.32}$ & 594 & 563 \\  
7.55-7.65 & cpl & -1.07$^{+0.03}_{-0.03}$ & - & 680.32$^{+51.91}_{-47.08}$ & 1.24$^{+6.10}_{-5.73}\times$10$^{-4}$ & 792.14$^{+46.38}_{-48.06}$ & 562 & 563 \\  
7.65-7.75 & cpl & -0.97$^{+0.03}_{-0.03}$ & - & 699.4$^{+43.64}_{-38.72}$ & 1.66$^{+7.73}_{-6.91}\times$10$^{-4}$ & 838.98$^{+40.62}_{-39.63}$ & 521 & 563 \\  
7.75-7.85 & cpl & -0.97$^{+0.03}_{-0.03}$ & - & 859.91$^{+53.66}_{-42.1}$ & 2.05$^{+9.20}_{-8.09}\times$10$^{-4}$ & 873.73$^{+41.06}_{-38.28}$ & 528 & 563 \\  
7.85-7.95 & cpl & -0.95$^{+0.02}_{-0.03}$ & - & 1087.21$^{+62.66}_{-57.74}$ & 2.53$^{+1.13}_{-1.04}\times$10$^{-4}$ & 836.14$^{+36.02}_{-33.6}$ & 552 & 563 \\  
7.95-8.05 & cpl & -0.82$^{+0.03}_{-0.03}$ & - & 1112.76$^{+57.25}_{-50.48}$ & 3.46$^{+1.34}_{-1.30}\times$10$^{-4}$ & 872.71$^{+28.8}_{-27.17}$ & 584 & 563 \\
8.05-8.15 & cpl & -0.77$^{+0.03}_{-0.02}$ & - & 1079.32$^{+43.81}_{-42.55}$ & 3.79$^{+1.28}_{-1.29}\times$10$^{-4}$ & 902.99$^{+26.01}_{-25.68}$ & 566 & 563 \\
8.15-8.25 & cpl & -0.77$^{+0.02}_{-0.03}$ & - & 1278.25$^{+48.33}_{-44.22}$ & 4.62$^{+1.61}_{-1.40}\times$10$^{-4}$ & 946.66$^{+27.19}_{-26.46}$ & 524 & 563 \\
8.25-8.35 & cpl & -0.77$^{+0.02}_{-0.02}$ & - & 1360.15$^{+52.8}_{-53.1}$ & 5.01$^{+1.76}_{-1.54}\times$10$^{-4}$ & 973.0$^{+25.63}_{-27.44}$ & 573 & 563 \\
8.35-8.45 & cpl & -0.79$^{+0.02}_{-0.02}$ & - & 1480.55$^{+61.9}_{-58.84}$ & 5.12$^{+1.76}_{-1.76}\times$10$^{-4}$ & 940.61$^{+26.48}_{-25.22}$ & 577 & 563 \\
8.45-8.55 & cpl & -0.72$^{+0.02}_{-0.02}$ & - & 1494.68$^{+55.02}_{-55.86}$ & 5.71$^{+2.00}_{-1.90}\times$10$^{-4}$ & 933.92$^{+23.48}_{-22.96}$ & 574 & 563 \\
8.55-8.65 & cpl & -0.73$^{+0.02}_{-0.02}$ & - & 1189.72$^{+47.35}_{-39.52}$ & 4.09$^{+1.51}_{-1.36}\times$10$^{-4}$ & 832.82$^{+22.79}_{-21.21}$ & 552 & 563 \\
8.65-8.75 & cpl & -0.76$^{+0.03}_{-0.03}$ & - & 923.42$^{+42.16}_{-40.19}$ & 2.57$^{+9.86}_{-9.85}\times$10$^{-4}$ & 702.11$^{+21.43}_{-23.47}$ & 514 & 563 \\
8.75-8.85 & cpl & -0.78$^{+0.03}_{-0.03}$ & - & 1013.73$^{+44.89}_{-41.23}$ & 2.81$^{+1.10}_{-1.06}\times$10$^{-4}$ & 718.52$^{+24.35}_{-22.64}$ & 480 & 563 \\
8.85-8.95 & cpl & -0.81$^{+0.03}_{-0.03}$ & - & 1054.25$^{+49.4}_{-45.52}$ & 2.67$^{+1.13}_{-1.03}\times$10$^{-4}$ & 696.95$^{+27.41}_{-23.19}$ & 526 & 563 \\
8.95-9.05 & cpl & -0.81$^{+0.03}_{-0.03}$ & - & 968.17$^{+54.06}_{-42.97}$ & 2.38$^{+1.04}_{-9.15}\times$10$^{-4}$ & 667.15$^{+24.98}_{-24.73}$ & 476 & 563 \\
9.05-9.15 & cpl & -0.85$^{+0.03}_{-0.03}$ & - & 899.17$^{+47.94}_{-41.87}$ & 2.06$^{+8.90}_{-8.60}\times$10$^{-4}$ & 668.04$^{+28.57}_{-27.13}$ & 538 & 563 \\
9.15-9.25 & cpl & -0.96$^{+0.03}_{-0.03}$ & - & 901.98$^{+50.1}_{-46.51}$ & 1.81$^{+7.77}_{-7.55}\times$10$^{-4}$ & 719.18$^{+35.82}_{-31.39}$ & 517 & 563 \\
9.25-9.35 & cpl & -0.89$^{+0.03}_{-0.03}$ & - & 936.61$^{+49.9}_{-45.19}$ & 2.36$^{+1.03}_{-9.32}\times$10$^{-4}$ & 798.22$^{+33.19}_{-32.34}$ & 601 & 563 \\
9.35-9.45 & cpl & -0.83$^{+0.03}_{-0.03}$ & - & 855.62$^{+38.97}_{-36.16}$ & 2.64$^{+9.65}_{-9.18}\times$10$^{-4}$ & 865.73$^{+33.87}_{-30.56}$ & 542 & 563 \\
9.45-9.55 & cpl & -0.88$^{+0.03}_{-0.03}$ & - & 887.19$^{+43.87}_{-41.31}$ & 2.59$^{+1.06}_{-9.32}\times$10$^{-4}$ & 903.06$^{+34.19}_{-32.92}$ & 567 & 563 \\
9.55-9.65 & cpl & -0.93$^{+0.03}_{-0.03}$ & - & 1244.0$^{+73.21}_{-67.32}$ & 2.82$^{+1.26}_{-1.18}\times$10$^{-4}$ & 802.19$^{+33.39}_{-31.64}$ & 626 & 563 \\
9.65-9.75 & cpl & -0.92$^{+0.03}_{-0.02}$ & - & 1491.59$^{+84.84}_{-81.84}$ & 2.78$^{+1.24}_{-1.21}\times$10$^{-4}$ & 662.19$^{+24.85}_{-23.41}$ & 575 & 563 \\
9.75-9.85 & cpl & -0.92$^{+0.02}_{-0.03}$ & - & 1567.51$^{+87.72}_{-83.27}$ & 2.92$^{+1.30}_{-1.32}\times$10$^{-4}$ & 669.47$^{+26.78}_{-23.89}$ & 457 & 563 \\
9.85-9.95 & cpl & -0.96$^{+0.03}_{-0.03}$ & - & 1552.87$^{+108.85}_{-90.65}$ & 2.81$^{+1.40}_{-1.24}\times$10$^{-4}$ & 699.96$^{+30.69}_{-29.25}$ & 521 & 563 \\
9.95-10.05 & cpl & -0.89$^{+0.03}_{-0.03}$ & - & 1352.87$^{+70.46}_{-70.24}$ & 2.82$^{+1.25}_{-1.18}\times$10$^{-4}$ & 688.84$^{+25.8}_{-27.07}$ & 569 & 563 \\
10.05-10.15 & cpl & -0.85$^{+0.02}_{-0.02}$ & - & 1320.74$^{+67.92}_{-58.74}$ & 3.60$^{+1.53}_{-1.37}\times$10$^{-4}$ & 827.03$^{+24.85}_{-27.61}$ & 465 & 479 \\
10.15-10.25 & cpl & -0.83$^{+0.02}_{-0.02}$ & - & 1227.45$^{+57.22}_{-50.18}$ & 3.79$^{+1.47}_{-1.39}\times$10$^{-4}$ & 894.3$^{+30.79}_{-28.2}$ & 449 & 479 \\
10.25-10.35 & cpl & -0.82$^{+0.03}_{-0.03}$ & - & 1116.27$^{+56.73}_{-48.54}$ & 3.33$^{+1.30}_{-1.26}\times$10$^{-4}$ & 833.95$^{+28.64}_{-23.78}$ & 512 & 479 \\
10.35-10.45 & cpl & -0.81$^{+0.03}_{-0.03}$ & - & 963.81$^{+58.42}_{-47.24}$ & 2.44$^{+1.14}_{-9.90}\times$10$^{-4}$ & 691.43$^{+26.87}_{-25.69}$ & 481 & 479 \\
10.45-10.55 & cpl & -0.84$^{+0.03}_{-0.03}$ & - & 951.07$^{+48.46}_{-47.39}$ & 2.30$^{+1.02}_{-9.73}\times$10$^{-4}$ & 702.89$^{+27.28}_{-26.7}$ & 451 & 479 \\
10.55-10.65 & cpl & -0.87$^{+0.03}_{-0.03}$ & - & 948.76$^{+46.24}_{-44.82}$ & 2.68$^{+1.06}_{-1.02}\times$10$^{-4}$ & 851.75$^{+31.42}_{-30.73}$ & 518 & 479 \\
10.65-10.75 & cpl & -0.97$^{+0.03}_{-0.03}$ & - & 997.18$^{+59.86}_{-55.16}$ & 2.38$^{+1.05}_{-9.77}\times$10$^{-4}$ & 892.11$^{+43.96}_{-38.88}$ & 492 & 479 \\
10.75-10.85 & cpl & -0.92$^{+0.03}_{-0.03}$ & - & 952.61$^{+58.42}_{-54.58}$ & 2.08$^{+1.05}_{-9.75}\times$10$^{-4}$ & 737.06$^{+34.38}_{-33.54}$ & 431 & 479 \\
10.85-10.95 & cpl & -0.93$^{+0.03}_{-0.03}$ & - & 896.41$^{+52.96}_{-49.62}$ & 1.95$^{+8.94}_{-8.81}\times$10$^{-4}$ & 732.7$^{+32.68}_{-33.68}$ & 464 & 479 \\
10.95-11.05 & cpl & $-0.99^{+0.03}_{-0.03}$ & - & $802.39^{+55.0}_{-50.08}$ & $1.50^{+7.54}_{-6.91}\times 10^{-4}$ & $708.57^{+37.94}_{-35.1}$ & 503 & 479 \\
11.05-11.15 & cpl & $-1.0^{+0.04}_{-0.04}$ & - & $654.9^{+52.39}_{-49.42}$ & $1.15^{+6.39}_{-6.13}\times 10^{-4}$ & $663.65^{+44.44}_{-37.78}$ & 504 & 479 \\
11.15-11.25 & cpl & $-1.03^{+0.03}_{-0.03}$ & - & $775.6^{+60.7}_{-51.22}$ & $1.59^{+7.81}_{-7.61}\times 10^{-4}$ & $835.52^{+47.18}_{-44.35}$ & 515 & 479 \\
11.25-11.35 & cpl & $-0.96^{+0.03}_{-0.03}$ & - & $838.51^{+50.26}_{-45.67}$ & $2.04^{+8.96}_{-8.80}\times 10^{-4}$ & $860.46^{+45.03}_{-37.53}$ & 503 & 479 \\
11.35-11.45 & cpl & $-1.02^{+0.03}_{-0.03}$ & - & $885.34^{+62.66}_{-55.04}$ & $1.77^{+9.02}_{-8.61}\times 10^{-4}$ & $822.21^{+45.25}_{-41.95}$ & 471 & 479 \\
11.45-11.55 & cpl & $-1.1^{+0.03}_{-0.03}$ & - & $834.46^{+82.3}_{-67.3}$ & $1.44^{+9.65}_{-8.28}\times 10^{-4}$ & $843.5^{+55.38}_{-50.49}$ & 502 & 479 \\
11.55-11.65 & cpl & $-1.12^{+0.04}_{-0.03}$ & - & $751.03^{+78.62}_{-67.63}$ & $1.27^{+8.29}_{-7.33}\times 10^{-4}$ & $848.28^{+54.62}_{-54.67}$ & 521 & 479 \\
11.65-11.75 & cpl & $-1.14^{+0.04}_{-0.04}$ & - & $691.63^{+76.55}_{-73.38}$ & $1.04^{+6.95}_{-7.00}\times 10^{-4}$ & $760.04^{+57.45}_{-50.79}$ & 501 & 479 \\
11.75-11.85 & cpl & $-1.17^{+0.03}_{-0.03}$ & - & $919.74^{+96.56}_{-81.36}$ & $1.35^{+7.92}_{-7.94}\times 10^{-4}$ & $862.77^{+60.45}_{-59.33}$ & 492 & 479 \\
11.85-11.95 & cpl & $-1.0^{+0.03}_{-0.03}$ & - & $1013.38^{+64.2}_{-65.35}$ & $2.17^{+1.01}_{-1.00}\times 10^{-4}$ & $860.05^{+45.04}_{-40.44}$ & 554 & 479 \\
11.95-12.05 & cpl & $-1.01^{+0.03}_{-0.02}$ & - & $1021.03^{+56.4}_{-57.94}$ & $2.27^{+9.96}_{-9.45}\times 10^{-4}$ & $914.1^{+40.25}_{-40.57}$ & 473 & 479 \\
12.05-12.15 & cpl & $-1.08^{+0.03}_{-0.03}$ & - & $1029.47^{+79.31}_{-75.37}$ & $1.73^{+9.54}_{-8.79}\times 10^{-4}$ & $813.82^{+47.75}_{-42.84}$ & 473 & 479 \\
12.15-12.25 & cpl & $-1.06^{+0.03}_{-0.03}$ & - & $979.55^{+83.82}_{-76.57}$ & $1.50^{+7.97}_{-7.93}\times 10^{-4}$ & $700.83^{+39.8}_{-38.06}$ & 488 & 479 \\
12.25-12.35 & cpl & $-1.06^{+0.03}_{-0.03}$ & - & $962.13^{+74.61}_{-65.89}$ & $1.59^{+7.98}_{-7.78}\times 10^{-4}$ & $751.44^{+42.39}_{-38.67}$ & 467 & 479 \\
12.35-12.45 & cpl & $-1.09^{+0.03}_{-0.03}$ & - & $1062.05^{+78.94}_{-74.27}$ & $1.78^{+9.98}_{-8.77}\times 10^{-4}$ & $844.6^{+48.44}_{-44.45}$ & 449 & 479 \\
12.45-12.55 & cpl & $-1.07^{+0.03}_{-0.03}$ & - & $1000.07^{+77.94}_{-65.22}$ & $1.68^{+8.95}_{-8.01}\times 10^{-4}$ & $784.15^{+50.91}_{-45.63}$ & 528 & 479 \\
12.55-12.65 & cpl & $-1.02^{+0.03}_{-0.03}$ & - & $953.95^{+66.62}_{-58.34}$ & $1.65^{+8.56}_{-8.11}\times 10^{-4}$ & $724.21^{+38.2}_{-36.56}$ & 500 & 479 \\
12.65-12.75 & cpl & $-1.04^{+0.03}_{-0.03}$ & - & $920.91^{+64.16}_{-65.01}$ & $1.72^{+8.95}_{-8.92}\times 10^{-4}$ & $803.63^{+48.41}_{-42.5}$ & 555 & 479 \\
12.75-12.85 & cpl & $-1.1^{+0.03}_{-0.03}$ & - & $882.7^{+67.65}_{-62.08}$ & $1.73^{+9.56}_{-7.87}\times 10^{-4}$ & $962.82^{+49.82}_{-53.02}$ & 521 & 479 \\
12.85-12.95 & cpl & $-1.01^{+0.03}_{-0.03}$ & - & $767.48^{+43.31}_{-42.21}$ & $2.05^{+8.61}_{-7.90}\times 10^{-4}$ & $1048.12^{+50.53}_{-48.43}$ & 545 & 479 \\
12.95-13.05 & cpl & $-1.01^{+0.03}_{-0.03}$ & - & $694.56^{+39.49}_{-34.75}$ & $2.05^{+8.13}_{-7.98}\times 10^{-4}$ & $1127.99^{+53.91}_{-50.38}$ & 523 & 479 \\
13.05-13.15 & cpl & $-1.09^{+0.02}_{-0.03}$ & - & $764.42^{+50.25}_{-43.37}$ & $1.92^{+9.08}_{-7.66}\times 10^{-4}$ & $1172.08^{+60.55}_{-55.41}$ & 514 & 479 \\
13.15-13.25 & cpl & $-1.07^{+0.03}_{-0.03}$ & - & $836.69^{+52.85}_{-46.49}$ & $2.14^{+8.87}_{-8.20}\times 10^{-4}$ & $1165.5^{+55.05}_{-52.93}$ & 577 & 479 \\
13.25-13.35 & cpl & $-1.11^{+0.03}_{-0.03}$ & - & $922.19^{+67.78}_{-61.81}$ & $2.20^{+1.04}_{-9.88}\times 10^{-4}$ & $1224.47^{+63.07}_{-59.3}$ & 551 & 479 \\
13.35-13.45 & cpl & $-1.14^{+0.03}_{-0.03}$ & - & $1017.63^{+81.52}_{-73.25}$ & $2.04^{+1.06}_{-9.98}\times 10^{-4}$ & $1136.46^{+62.8}_{-60.62}$ & 521 & 479 \\
13.45-13.55 & cpl & $-1.14^{+0.04}_{-0.03}$ & - & $935.38^{+100.3}_{-79.43}$ & $1.54^{+9.82}_{-8.59}\times 10^{-4}$ & $909.31^{+62.24}_{-63.74}$ & 495 & 479 \\
13.55-13.65 & cpl & $-1.2^{+0.04}_{-0.04}$ & - & $834.26^{+113.42}_{-99.19}$ & $1.17^{+8.73}_{-8.22}\times 10^{-4}$ & $880.99^{+75.43}_{-63.57}$ & 483 & 479 \\
13.65-13.75 & cpl & $-1.26^{+0.04}_{-0.03}$ & - & $1002.12^{+138.23}_{-120.26}$ & $1.23^{+8.73}_{-8.85}\times 10^{-4}$ & $953.27^{+73.25}_{-72.15}$ & 508 & 479 \\
13.75-13.85 & cpl & $-1.25^{+0.04}_{-0.04}$ & - & $842.54^{+127.15}_{-101.66}$ & $1.09^{+8.61}_{-7.80}\times 10^{-4}$ & $936.84^{+72.99}_{-75.86}$ & 532 & 479 \\
13.85-13.95 & cpl & $-1.2^{+0.03}_{-0.04}$ & - & $781.26^{+69.56}_{-67.71}$ & $1.12^{+6.28}_{-6.03}\times 10^{-4}$ & $875.12^{+73.82}_{-61.32}$ & 507 & 479 \\
13.95-14.05 & cpl & -1.22$^{+0.03}_{-0.03}$ & - & 844.48$^{+83.95}_{-79.87}$ & 1.27$^{+7.39}_{-7.00}\times$10$^{-4}$ & 992.39$^{+79.98}_{-65.85}$ & 520 & 479 \\ 
14.05-14.15 & cpl & $-1.22^{+0.03}_{-0.03}$ & - & $878.23^{+74.58}_{-71.92}$ & $1.33^{+7.03}_{-6.80}\times 10^{-4}$ & $999.97^{+71.46}_{-63.06}$ & 542 & 479 \\
14.15-14.25 & cpl & $-1.18^{+0.03}_{-0.03}$ & - & $1078.32^{+99.66}_{-86.3}$ & $1.59^{+8.98}_{-8.42}\times 10^{-4}$ & $941.13^{+55.53}_{-54.87}$ & 443 & 479 \\
14.25-14.35 & cpl & $-1.12^{+0.03}_{-0.03}$ & - & $1079.29^{+79.79}_{-72.67}$ & $1.79^{+8.41}_{-8.23}\times 10^{-4}$ & $902.16^{+52.59}_{-45.38}$ & 497 & 479 \\
14.35-14.45 & cpl & $-1.21^{+0.03}_{-0.03}$ & - & $995.37^{+95.64}_{-98.14}$ & $1.30^{+7.35}_{-7.84}\times 10^{-4}$ & $867.57^{+67.18}_{-60.7}$ & 491 & 479 \\
14.45-14.55 & cpl & $-1.29^{+0.05}_{-0.04}$ & - & $493.47^{+81.49}_{-59.96}$ & $7.39^{+6.17}_{-5.17}\times 10^{-5}$ & $1000.46^{+101.15}_{-90.37}$ & 546 & 479 \\
14.55-14.65 & cpl & $-1.27^{+0.05}_{-0.04}$ & - & $549.12^{+71.26}_{-59.16}$ & $9.12^{+6.20}_{-5.56}\times 10^{-5}$ & $1097.6^{+97.77}_{-101.69}$ & 555 & 479 \\
14.65-14.75 & cpl & $-1.37^{+0.03}_{-0.03}$ & - & $715.65^{+94.06}_{-78.48}$ & $1.02^{+6.31}_{-6.37}\times 10^{-4}$ & $1344.12^{+113.86}_{-98.57}$ & 520 & 479 \\
14.75-14.85 & cpl & $-1.41^{+0.03}_{-0.03}$ & - & $892.66^{+148.52}_{-113.1}$ & $1.11^{+8.48}_{-7.24}\times 10^{-4}$ & $1430.72^{+123.7}_{-111.67}$ & 560 & 479 \\
14.85-14.95 & cpl & $-1.38^{+0.02}_{-0.03}$ & - & $1053.25^{+157.08}_{-129.84}$ & $1.32^{+9.17}_{-8.61}\times 10^{-4}$ & $1402.14^{+118.64}_{-94.51}$ & 512 & 479 \\
14.95-15.05 & cpl & $-1.19^{+0.03}_{-0.03}$ & - & $700.87^{+65.31}_{-61.19}$ & $1.28^{+6.72}_{-6.93}\times 10^{-4}$ & $1056.83^{+70.65}_{-69.31}$ & 458 & 479 \\
15.05-15.55 & cpl & $-1.09^{+0.02}_{-0.02}$ & - & $457.92^{+29.96}_{-27.19}$ & $8.52^{+3.44}_{-3.25}\times 10^{-5}$ & $770.81^{+30.94}_{-27.63}$ & 609 & 479 \\
15.55-16.05 & cpl & $-1.19^{+0.02}_{-0.02}$ & - & $678.97^{+47.05}_{-43.61}$ & $8.94^{+3.67}_{-3.58}\times 10^{-5}$ & $750.8^{+30.38}_{-28.15}$ & 630 & 479 \\
16.05-16.55 & cpl & $-1.14^{+0.02}_{-0.02}$ & - & $739.1^{+51.41}_{-45.76}$ & $8.66^{+3.70}_{-3.60}\times 10^{-5}$ & $608.59^{+25.05}_{-23.74}$ & 584 & 479 \\
16.55-17.05 & cpl & $-1.15^{+0.03}_{-0.03}$ & - & $439.02^{+44.19}_{-39.57}$ & $5.48^{+3.26}_{-2.97}\times 10^{-5}$ & $588.64^{+32.28}_{-33.13}$ & 654 & 479 \\
17.05-17.55 & cpl & $-1.24^{+0.03}_{-0.03}$ & - & $520.47^{+56.93}_{-48.69}$ & $5.38^{+3.35}_{-2.93}\times 10^{-5}$ & $617.04^{+34.44}_{-32.27}$ & 563 & 479 \\
17.55-18.05 & cpl & $-1.28^{+0.03}_{-0.02}$ & - & $322.7^{+17.3}_{-15.26}$ & $2.37^{+7.29}_{-8.18}\times 10^{-5}$ & $413.09^{+21.51}_{-23.63}$ & 634 & 601 \\
18.05-18.55 & cpl & $-1.58^{+0.03}_{-0.04}$ & - & $238.52^{+30.7}_{-24.76}$ & $1.04^{+6.36}_{-6.02}\times 10^{-5}$ & $393.8^{+39.85}_{-30.66}$ & 665 & 601 \\
18.55-19.05 & cpl & $-1.38^{+0.02}_{-0.02}$ & - & $356.51^{+20.61}_{-18.79}$ & $2.35^{+9.00}_{-8.77}\times 10^{-5}$ & $477.34^{+24.74}_{-26.15}$ & 846 & 601 \\
19.05-19.55 & cpl & $-1.26^{+0.01}_{-0.01}$ & - & $501.42^{+19.88}_{-15.65}$ & $5.84^{+1.62}_{-1.43}\times 10^{-5}$ & $722.36^{+25.16}_{-22.2}$ & 965 & 601 \\
19.55-20.05 & cpl & $-1.24^{+0.01}_{-0.01}$ & - & $680.58^{+19.76}_{-19.99}$ & $7.74^{+1.73}_{-1.75}\times 10^{-5}$ & $743.55^{+20.87}_{-20.48}$ & 880 & 601 \\
20.05-20.55 & cpl & $-1.2^{+0.01}_{-0.01}$ & - & $812.2^{+24.88}_{-23.34}$ & $8.15^{+1.96}_{-1.82}\times 10^{-5}$ & $626.7^{+18.39}_{-17.1}$ & 893 & 601 \\
20.55-21.05 & cpl & $-1.16^{+0.01}_{-0.02}$ & - & $600.57^{+18.32}_{-16.67}$ & $5.62^{+1.46}_{-1.37}\times 10^{-5}$ & $491.63^{+17.43}_{-16.31}$ & 800 & 601 \\
21.05-21.55 & cpl & $-1.14^{+0.01}_{-0.01}$ & - & $601.31^{+15.09}_{-15.06}$ & $7.53^{+1.79}_{-1.86}\times 10^{-5}$ & $618.85^{+18.75}_{-17.75}$ & 813 & 601 \\
21.55-22.05 & cpl & $-1.18^{+0.01}_{-0.01}$ & - & $594.88^{+18.98}_{-18.76}$ & $6.14^{+1.56}_{-1.57}\times 10^{-5}$ & $558.23^{+17.79}_{-17.61}$ & 760 & 601 \\
22.05-22.55 & cpl & $-1.23^{+0.01}_{-0.01}$ & - & $583.11^{+17.53}_{-19.12}$ & $5.10^{+1.45}_{-1.38}\times 10^{-5}$ & $527.29^{+16.07}_{-16.12}$ & 914 & 601 \\
22.55-23.05 & cpl & $-1.2^{+0.02}_{-0.02}$ & - & $451.57^{+16.13}_{-13.87}$ & $4.36^{+1.18}_{-1.05}\times 10^{-5}$ & $501.17^{+18.94}_{-17.65}$ & 743 & 601 \\
23.05-24.55 & cpl & $-1.26^{+0.01}_{-0.01}$ & - & $399.74^{+8.97}_{-8.8}$ & $3.94^{+7.24}_{-6.66}\times 10^{-5}$ & $572.43^{+14.72}_{-13.05}$ & 1280 & 601 \\
24.55-26.05 & cpl & $-1.3^{+0.01}_{-0.01}$ & - & $401.48^{+8.82}_{-8.46}$ & $4.28^{+6.95}_{-6.68}\times 10^{-5}$ & $673.9^{+13.94}_{-14.92}$ & 1297 & 601 \\
26.05-27.55 & cpl & $-1.5^{+0.01}_{-0.01}$ & - & $230.16^{+7.57}_{-6.51}$ & $2.27^{+4.31}_{-4.31}\times 10^{-5}$ & $739.3^{+23.67}_{-23.5}$ & 896 & 322 \\
27.55-29.05 & cpl & $-1.44^{+0.01}_{-0.01}$ & - & $525.4^{+17.31}_{-16.3}$ & $3.42^{+7.21}_{-7.11}\times 10^{-5}$ & $651.05^{+18.2}_{-17.79}$ & 893 & 322 \\
29.05-30.55 & cpl & $-1.44^{+0.01}_{-0.01}$ & - & $468.84^{+22.79}_{-21.69}$ & $1.99^{+5.75}_{-5.33}\times 10^{-5}$ & $399.25^{+14.57}_{-13.01}$ & 569 & 322 \\
30.55-32.05 & cpl & $-1.42^{+0.02}_{-0.02}$ & - & $170.74^{+7.17}_{-6.79}$ & $9.91^{+2.60}_{-2.99}\times 10^{-6}$ & $330.19^{+14.63}_{-14.52}$ & 515 & 322 \\
32.05-34.7 & cpl & $-1.54^{+0.01}_{-0.01}$ & - & $202.73^{+7.15}_{-7.22}$ & $1.19^{+2.59}_{-2.57}\times 10^{-5}$ & $448.82^{+16.66}_{-16.36}$ & 751 & 322 \\
34.7-36.2 & cpl & $-1.47^{+0.02}_{-0.01}$ & - & $266.89^{+12.11}_{-10.48}$ & $1.42^{+4.24}_{-3.75}\times 10^{-5}$ & $411.16^{+17.13}_{-17.06}$ & 548 & 322 \\
36.2-39.0 & cpl & $-1.47^{+0.01}_{-0.02}$ & - & $183.48^{+6.07}_{-5.74}$ & $9.56^{+2.27}_{-2.29}\times 10^{-6}$ & $332.05^{+12.8}_{-12.5}$ & 524 & 322 \\
39.0-42.8 & cpl & $-1.63^{+0.02}_{-0.02}$ & - & $121.89^{+5.76}_{-5.07}$ & $5.19^{+1.57}_{-1.47}\times 10^{-6}$ & $282.28^{+13.88}_{-12.51}$ & 488 & 322 \\
42.8-47.7 & cpl & $-1.66^{+0.02}_{-0.02}$ & - & $150.29^{+6.85}_{-6.16}$ & $5.24^{+1.51}_{-1.48}\times 10^{-6}$ & $274.07^{+12.15}_{-11.44}$ & 533 & 322 \\
47.7-53.0 & cpl & $-1.55^{+0.02}_{-0.02}$ & - & $122.27^{+5.32}_{-4.63}$ & $3.77^{+1.10}_{-1.12}\times 10^{-6}$ & $180.86^{+9.26}_{-8.94}$ & 517 & 322 \\
53.0-58.0 & cpl & $-1.61^{+0.03}_{-0.02}$ & - & $79.24^{+3.82}_{-3.54}$ & $2.46^{+1.12}_{-1.01}\times 10^{-6}$ & $154.34^{+10.77}_{-10.21}$ & 474 & 322 \\
58.0-63.0 & cpl & $-1.63^{+0.03}_{-0.03}$ & - & $89.92^{+4.62}_{-4.48}$ & $2.49^{+1.12}_{-1.07}\times 10^{-6}$ & $153.46^{+11.54}_{-10.41}$ & 382 & 322 \\
63.0-68.0 & cpl & $-1.72^{+0.05}_{-0.04}$ & - & $96.8^{+11.6}_{-9.2}$ & $1.99^{+1.39}_{-1.36}\times 10^{-6}$ & $133.71^{+13.93}_{-14.82}$ & 410 & 322 \\
68.0-73.0 & cpl & $-1.62^{+0.05}_{-0.04}$ & - & $115.63^{+14.02}_{-11.13}$ & $1.59^{+1.10}_{-1.11}\times 10^{-6}$ & $86.8^{+8.9}_{-9.45}$ & 381 & 322 \\
73.0-78.0 & cpl & $-1.64^{+0.09}_{-0.08}$ & - & $37.84^{+5.57}_{-5.29}$ & $4.38^{+7.94}_{-7.41}\times 10^{-7}$ & $37.91^{+8.63}_{-8.51}$ & 329 & 322 \\
78.0-83.0 & cpl & $-1.69^{+0.07}_{-0.07}$ & - & $76.4^{+8.7}_{-7.86}$ & $8.71^{+1.07}_{-8.65}\times 10^{-7}$ & $60.76^{+12.55}_{-9.34}$ & 353 & 322 \\
83.0-88.0 & cpl & $-1.66^{+0.07}_{-0.07}$ & - & $66.7^{+10.2}_{-7.85}$ & $6.54^{+7.30}_{-6.14}\times 10^{-7}$ & $46.23^{+8.04}_{-6.84}$ & 372 & 322 \\
88.0-93.0 & cpl & $-1.72^{+0.07}_{-0.06}$ & - & $49.54^{+7.66}_{-6.67}$ & $6.82^{+8.63}_{-8.40}\times 10^{-7}$ & $56.97^{+9.79}_{-10.13}$ & 319 & 322 \\
93.0-98.0 & pl & $-1.94^{+0.04}_{-0.04}$ & - & - & $7.01^{+1.29}_{-1.28}\times 10^{-7}$ & $39.74^{+10.0}_{-8.46}$ & 320 & 322 \\
98.0-103.0 & pl & $-1.94^{+0.07}_{-0.08}$ & - & - & $5.24^{+1.54}_{-1.56}\times 10^{-7}$ & $30.1^{+13.78}_{-10.88}$ & 344 & 322 \\
103.0-113.0 & pl & $-2.17^{+0.08}_{-0.09}$ & - & - & $3.56^{+1.08}_{-1.06}\times 10^{-7}$ & $41.68^{+21.55}_{-14.75}$ & 296 & 322 \\
\botrule
\caption{\textcolor{black}{Spectral fitting results: This table summarizes the best-fit parameters for each time interval of GRB 230307A. The columns detail the time interval, the selected best-fit model, and the $\alpha$ parameter for the PL, CPL, or BAND functions, where the corresponding model represnts the best model fit. For the BAND function, the $\beta$ parameter is included in intervals where it provides the best fit. Also listed are the peak energy, flux, photon flux, CSTAT value, and degrees of freedom (DOF), which together evaluate the goodness of fit for the chosen model. Our analysis also considered the inclusion of an additional spectral component, such as a thermal component, but no statistically significant evidence was found to justify its necessity.}}\label{tab:fit_result}\\
\end{longtable} 
\end{ThreePartTable}
\endgroup

\end{CJK*}
\end{document}